\documentclass[%
reprint,
superscriptaddress,
 amsmath,amssymb,
prl,
]{revtex4-2}
\usepackage{amssymb}
\usepackage{amsmath}
\usepackage{bbm}
\usepackage{bm}
\usepackage{epsfig}
\usepackage{color}
\usepackage{multirow}
\usepackage{graphicx}
\usepackage{booktabs}
\usepackage{dsfont}
\usepackage{tikz}
\usepackage{appendix}
\usepackage{makecell}
\usepackage{subfigure}
\usepackage{braket}
\usepackage{xcolor}
\usepackage[colorlinks,linkcolor=blue,anchorcolor=blue,citecolor=blue,urlcolor=blue]{hyperref}
\begin{document}

\title{Unified Bulk-Entanglement Correspondence in Non-Hermitian Systems}

\author{Xudong Zhang}
\affiliation{Department of Physics, Wuhan University of Technology, Wuhan 430070, China}

\author{Zhaoyu Sun}
\affiliation{School of Electrical and Electronic Engineering, Wuhan Polytechnic University, Wuhan 430023, China}

\author{Bin Guo}
\email{binguo@whut.edu.cn}
\affiliation{Department of Physics, Wuhan University of Technology, Wuhan 430070, China}

\date{\today} 

\begin{abstract}
The non-Hermitian skin effect (NHSE) fundamentally invalidates the conventional bulk-boundary correspondence (BBC), leading topological diagnostics into a crisis. While the non-Bloch polarization $P_{\beta}$ defined on the generalized Brillouin zone restores momentum-space topology, a direct, robust real-space bulk probe has remained elusive. We resolve this by establishing a universal correspondence between $P_{\beta}$ and the entanglement polarization $\chi$ of the biorthogonal ground state. Introducing a quasi-reciprocal Hamiltonian $\tilde{H}$ that removes the NHSE while preserving bulk topology, we rigorously prove the fundamental identity $P_{\beta} \equiv \chi(\tilde{H})\pmod 1$ in the thermodynamic limit under the quasi-locality assumption. Crucially, we demonstrate that this equivalence transcends the locality constraints that limit traditional topological invariants. While the conventional Resta polarization fails when $\tilde{H}$ becomes non-local due to the divergence of position variance, we reveal that $\chi(\tilde{H})$ remains robustly quantized, protected by the Fredholm index of Toeplitz operators. Our work thus identifies entanglement as the unique real-space diagnostic capable of capturing non-Bloch topology beyond the breakdown of locality, successfully restoring the BBC across diverse non-Hermitian systems such as line-gap, point-gap, and gapless phases, thereby unifying the geometric and entanglement paradigms in non-Hermitian physics.
\end{abstract}

\maketitle

{\it Introduction.}---
Non-Hermitian physics has emerged as a frontier in condensed matter and quantum optics, offering a powerful framework for describing open quantum systems with gain and loss~\cite{Bender1998, Ashida2020, ElGanainy2018, Gong2018, Bergholtz2021, Okuma2023, Wang2021, Bender2007}. Distinct from conventional Hermitian systems, non-Hermitian Hamiltonians admit complex energy spectra and exhibit exotic phenomena such as exceptional points (EPs)~\cite{Lu2014, Miri2019, Heiss2012, CanosValero2025,Yokomizo2020, Zhen2015, Doppler2016, Budich2019, Kawabata2021, Xiong2021, Luitz2019, Cai2022, Hu2017, Yang2021, Mandal2021, Xiao2024a, Zhu2018} and the non-Hermitian skin effect (NHSE)~\cite{Yao2018, Lee2019, Okuma2020, Lee2016, Longhi2020, Zhang2022a, Zhang2024a, Hou2024, Gliozzi2024, Okugawa2020, Yokomizo2021, Longhi2022, Song2019, Li2020, Sun2021b, Li2023e, Liu2024a, Shimomura2024, Kawabata2023, Zhou2025, Xu2025, Wang2023a}. The NHSE, in particular, is remarkable: under open boundary conditions (OBCs), a macroscopic number of bulk eigenstates accumulate exponentially at the system’s edge. This boundary localization invalidates conventional Bloch band topology and fundamentally breaks the bulk-boundary correspondence (BBC)~\cite{Kunst2018, Xiao2020, Trifunovic2019, Nakamura2024, Yang2020, Rhim2017, Tamura2021, Zhang2020a}, necessitating a transition from standard Bloch band theory to the non-Bloch framework, a shift that often entails the loss of real-space locality and thus challenges the applicability of traditional topological markers.

A major breakthrough came with the non-Bloch band theory~\cite{Yao2018, Yokomizo2019, Kawabata2019, Borgnia2020, Kaneshiro2025, Yokomizo2024, Zhang2020a, Yang2020, Yang2024, Wang2024b, Longhi2020a, Verma2024, Xue2021, Kawabata2020}, which restores the topological correspondence by introducing the generalized Brillouin zone (GBZ) contour $C_\beta$. The non-Bloch band theory extends the Bloch momentum $k$ into the complex plane $\beta=e^{ik}$, defining the non-Bloch Hamiltonian $H(\beta)$ whose eigenstates ${\vert u_R(\beta)\rangle, \vert u_L(\beta)\rangle}$ satisfy the biorthogonal relations $H(\beta)\vert u_R(\beta)\rangle = E(\beta)\vert u_R(\beta)\rangle$ and $\langle u_L(\beta)\vert H(\beta) = E(\beta)\langle u_L(\beta)\vert$, with normalization $\langle u_L(\beta)\vert u_R(\beta)\rangle = 1$. The resulting non-Bloch polarization $P_\beta$ is defined via the biorthogonal Berry connection $\mathcal{A}(\beta) = \langle u_L(\beta) \vert \partial_\beta u_R(\beta)\rangle$ integrated over the GBZ contour~\cite{Lee2020, Yang2024}
\begin{align}
	P_\beta = \frac{1}{2\pi i} \oint_{C_\beta} d\beta  \mathcal{A}(\beta) \pmod 1.
\end{align}

This polarization represents the topological phase (winding) of the  non-Bloch Wilson loop~\cite{Masuda2022}, $W_\beta = \exp(i 2\pi P_\beta) = \exp\left( \oint_{C_\beta} d\beta \mathcal{A}(\beta) \right)$. This quantity successfully restores the BBC by correctly predicting the topological edge states. However, $P_\beta$ remains an abstract momentum-space quantity defined on the complex GBZ. This limitation is critical because, in conventional Hermitian topological systems, momentum-space topology (such as the Zak phase) is strictly equivalent to bulk topological indices, establishing a fundamental identity between the real-space Resta polarization ($P(H)$) and the entanglement polarization ($\chi(H)$)~\cite{KingSmith1993, Resta1994, Watanabe2018, Zak1989, Xiao2010, Li2008, Alexandradinata2011}. This bulk-entanglement correspondence is a cornerstone of Hermitian topological physics. It is therefore a central and pressing question to identify the real-space entanglement invariant that corresponds to $P_\beta$ in non-Hermitian systems. This search, however, faces two fundamental obstacles: (i) Prior studies of the entanglement spectrum of the physical Hamiltonian $H$ have revealed its pathological behavior in the point-gapped regime. Due to the NHSE, the periodic boundary conditions-based entanglement spectrum becomes disconnected from OBC physics~\cite{Loic2019}. Consequently, the entanglement polarization $\chi(H)$, computed from this pathological spectrum, fundamentally fails to capture topological transitions~\cite{OrtegaTaberner2022}. (ii) Constructing a surrogate NHSE-free Hamiltonian $\tilde{H}$ from the GBZ appears promising, but it generically yields non-local hopping~\cite{Lee2020, Yang2024}; this non-locality violates the nearsightedness principle and causes the variance of the position operator to diverge, rendering the conventional Resta polarization ill-defined. 

Therefore, a central open question remains: {\emph {Can the momentum-space GBZ invariant $P_\beta$ be rigorously and quantitatively connected to a real-space entanglement invariant $\chi$?} In this Letter, we establish an exact topological equivalence between the non-Bloch polarization $P_\beta$ and the real-space entanglement polarization $\chi$, mediated by the quasi-reciprocal Hamiltonian $\tilde{H}$. Crucially, we identify that this correspondence is protected by the Index Theory of Toeplitz operators~\cite{Bottcher2006}. This equivalence, $P_{\beta} \equiv \chi(\tilde{H})\pmod 1$, persists even when $\tilde{H}$ becomes non-local due to a deformed GBZ. Unlike the Resta polarization which fails due to the divergence of the position operator, $\chi(\tilde{H})$ relies solely on the spectral gap of the correlation matrix, thus revealing a topological identity that transcends the strict locality constraints. This discovery establishes $\chi(\tilde{H})$ as a robust real-space diagnostic immune to the breakdown of locality, unifying the geometric (GBZ) and entanglement (EOS) paradigms into a single topological framework.

{\it Non-Bloch–Entanglement Correspondence: Topology Beyond Locality.}---
The construction of $\tilde{H}$ is equivalent to applying a complex momentum deformation, $\beta = e^{ik}e^{-\kappa(k)}$, to “untwist” the NHSE~\cite{Lee2020}. In simple models (e.g., those with only nearest-neighbor nonreciprocity), $\kappa(k)$ is a constant, the GBZ contour $C_\beta$ forms a simple circle, and $\tilde{H}$ remains local. However, in more generic systems with multiple nonreciprocal coupling ranges, such as the $t_3 \neq 0$ case analyzed in Fig.~\ref{Fig1}(c), canceling the multi-scale pumping requires $\kappa(k)$ to develop a nontrivial $k$-dependence. This $k$-dependence renders the GBZ contour $C_\beta$ noncircular, leading to power-law decay of the effective hopping terms in $\tilde{H}$ rather than the exponential localization required by the nearsightedness principle.

To investigate this regime, we formally define the surrogate Hamiltonian $\tilde{H}$ through an inverse generalized Fourier transform over the potentially nonanalytic contour $C_\beta$~\cite{Lee2020, Yang2024}
\begin{align}
	\tilde H = \sum_{j,l} \tilde t_{j-l}\, c_j^\dagger c_l ,\qquad\tilde t_n = \oint_{C_\beta} \frac{d\beta}{2\pi i \beta}\, H(\beta)\, \beta^{-n}.
\end{align}

\begin{table}[t]
    \caption{Summary of topological invariants and their validity.The table contrasts the breakdown of the conventional real-space Resta polarization $P(\tilde{H})$ against the robustness of the entanglement polarization $\chi(\tilde{H})$ in the non-Bloch framework.}
    \label{tab:summary}
    \centering
    \small
    \begin{tabular}{@{}lccc@{}}
        \hline\hline
        Framework & \makecell{Momentum \\ Space} & \makecell{Real \\ Space} & \makecell{Entanglement \\ Space} \\
        \hline
        Hermitian & $P$ (Bloch) & $P(H)$ (Resta) & $\chi(H)$ \\[2pt]
        Effective Construct & $H(\beta)$ & $\tilde{H}$ & $\tilde{P}$ \\[2pt]
        Non-Hermitian & \makecell{$P_\beta$ \\ (Non-Bloch)} & \makecell{$P(\tilde{H})$ \\ (\emph{Ill-defined})} & \makecell{$\chi(\tilde{H})$ \\ (\textbf{Robust})} \\
        \hline\hline
    \end{tabular}
\end{table}

This breakdown of locality causes the position variance to diverge, posing a central theoretical challenge for conventional real-space invariants. To frame our contribution, we summarize in Table~\ref{tab:summary} the relationships among topological invariants in distinct representations. This table provides a concise mapping between the momentum-space, real-space, and entanglement-space characterizations of topology, highlighting how their equivalence in Hermitian systems becomes subtle or even breaks down in the non-Hermitian regime. Specifically, it contrasts the Bloch polarization $P$, the Resta polarization $P(H)$, and the entanglement polarization $\chi(H)$ with their non-Hermitian counterparts $P_\beta$, $P(\tilde{H})$, and $\chi(\tilde{H})$, respectively.

As summarized in Table~\ref{tab:summary}, one might expect this Hermitian “triple equivalence” to naturally extend to the surrogate Hamiltonian $\tilde{H}$, which successfully restores the bulk-boundary correspondence. However, the intrinsic non-locality of $\tilde{H}$ fundamentally disrupts this unity, acting as a discriminator between different real-space invariants. To demonstrate this, we explicitly contrast two real-space formulations:

\begin{enumerate}
    \item The Resta polarization, which encodes topology geometrically via the expectation value of the position operator twist $\hat{X}$~\cite{Resta1998, Lee2020}
    \begin{align}
    	e^{i2\pi P(\tilde H)} = \langle \tilde\Psi_L\vert e^{\frac{2\pi i}{L}\hat X}\vert\tilde\Psi_R\rangle.
    \end{align}    
    \item The Entanglement polarization $\chi(\tilde{H})$, which is defined algebraically through the single-particle non-Bloch projector $\tilde{P} = \sum_{\text{occ}} |u_R^\sim \rangle \langle u_L^\sim |$. To compute $\chi(\tilde{H})$, one first restricts $\tilde{P}$ to a subsystem $A$, forming the correlation matrix $C^A_{ij} = \langle i | \tilde{P} | j \rangle, \quad i,j \in A$. The eigenvalues $\{\xi_\mu\}$ of $C^A$ constitute the entanglement spectrum. The entanglement polarization is then obtained by summing the eigenvalues over the subset $\mathcal{L}$ of modes that are localized at the left boundary of subsystem $A$~\cite{OrtegaTaberner2021, OrtegaTaberner2022}
    \begin{align}\label{Eq.4}
       \chi(\tilde H) = \sum_{\mu \in L} \xi_\mu \pmod 1.
    \end{align}
\end{enumerate}

While these two quantities coincide for local Hamiltonians, the nonlocality of $\tilde{H}$ fundamentally breaks this equivalence. The definition of Resta polarization $P(\tilde{H})$ relies on the nearsightedness principle, which is violated by the power-law hoppings in $\tilde{H}$ that cause the position variance to diverge, rendering $P(\tilde{H})$ ill-defined~\cite{Resta1994, Kohn1996}. As emphasized in Table~\ref{tab:summary}, this breakdown marks the key distinction between Hermitian and non-Hermitian topology: whereas $P(\tilde{H})$ loses its quantization due to nonlocality, $\chi(\tilde{H})$ remains a robust and well-defined topological indicator. The central question therefore arises: {\emph {Can the entanglement polarization $\chi(\tilde{H})$, which depends only on the single-particle projector and the presence of an entanglement gap, remain well-defined and quantized in the presence of nonlocality?}}

We rigorously prove that $P_{\beta} \equiv \chi(\tilde{H})\pmod 1$ when $\tilde{H}$ satisfies the quasi-locality condition (exponentially decaying hopping terms). Crucially, our analysis reveals that this equivalence extends beyond the quasi-local regime. Extensive numerical evidence demonstrates that $\chi(\tilde{H})$ remains quantized even when $\tilde{H}$ exhibits power-law decaying hoppings. This robustness establishes $\chi(\tilde{H})$ as the rigorous real-space dual to the non-Bloch topology, confirming the fundamental equivalence:
\begin{align}
	P_\beta = \chi(\tilde H)\pmod 1.
\end{align}

This discovery demonstrates that $\chi(\tilde{H})$ serves as the true and resilient real-space counterpart to the non-Bloch polarization $P_\beta$, succeeding precisely where the conventional Resta polarization $P(\tilde{H})$ fails. As detailed in Supplementary Material Section~\textcolor{blue}{\hyperref[sec: S3]{S3}}, this robustness stems from the Index Theory of Toeplitz operators~\cite{Bottcher2006}: unlike the Resta polarization which requires finite position variance, the quantization of $\chi(\tilde{H})$ is protected by the Fredholm index, which remains well-defined as long as the effective hoppings decay algebraically ($\alpha > 1$). This mechanism ensures the equivalence persists across generic non-Hermitian topological insulators, even in regimes dominated by a strong NHSE. Detailed derivations and the Toeplitz index analysis are provided in Supplementary Material Sections~\textcolor{blue}{\hyperref[sec: S1]{S1}},~\textcolor{blue}{\hyperref[sec: S2]{S2}} and \textcolor{blue}{\hyperref[sec: S3]{S3}}.

{\it Numerical Verification and Analysis.}---
We consider a prototypical one-dimensional non-Hermitian free-fermion system of the Hatano-Nelson type~\cite{Hatano1996, Yao2018, Yokomizo2019} (see Supplementary Material Section~\textcolor{blue}{\hyperref[sec: S5]{S5}} for details). While this specific model serves as a concrete example, our theoretical framework and main conclusions should be applicable to general one-dimensional non-Hermitian free-fermion systems with appropriate symmetries. Our discussion focuses on the general case where the GBZ is non-circular (e.g., $t_3 \neq 0$). Crucially, this deformation induces power-law decaying hoppings in $\tilde{H}$, creating the regime where the nearsightedness principle is violated and the Resta polarization fails. Validation in the local limit (circular GBZ, e.g., $t_3 = 0$) is presented in Supplementary Material Section~\textcolor{blue}{\hyperref[sec: S5]{S5}}, providing a foundational verification of our approach.

\begin{figure}[t]
\centering
\includegraphics[width=0.4\textwidth,keepaspectratio]{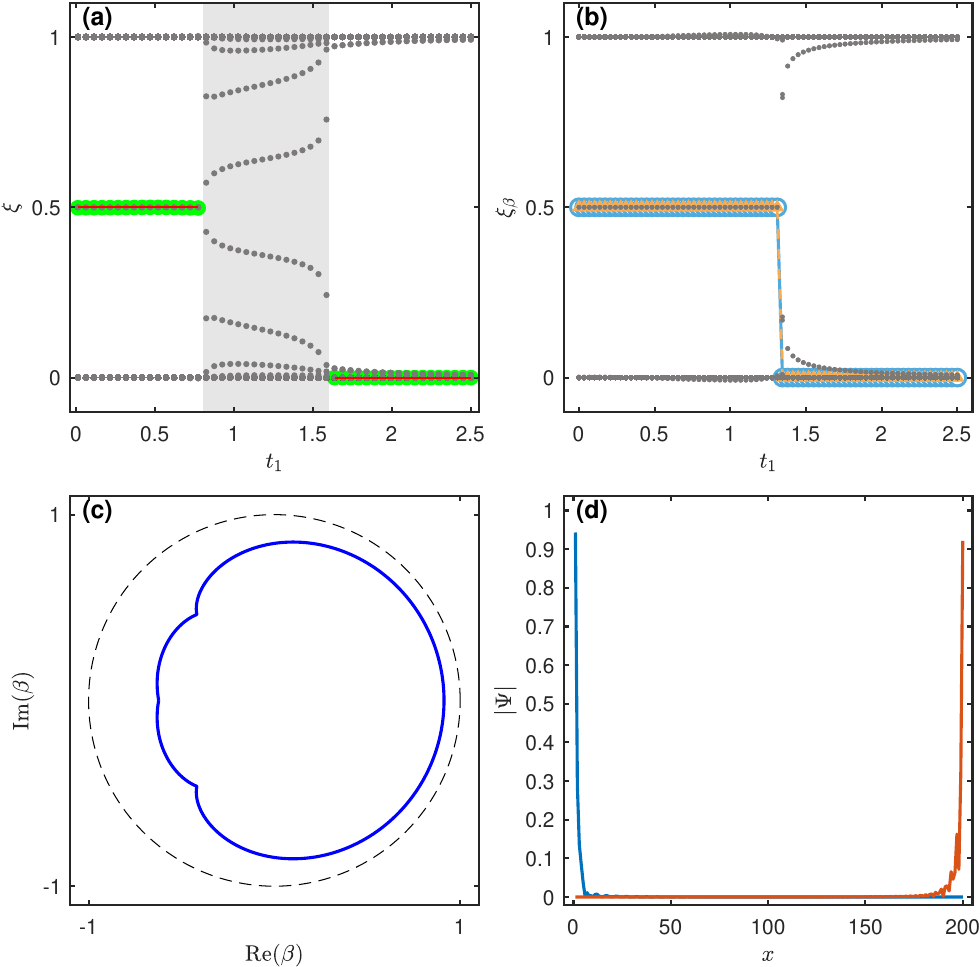}
\caption{Unifying bulk-boundary correspondence via entanglement. (a) Evolution of the periodic boundary conditions entanglement spectrum (grey) with $t_1$. The standard PBC polarization $\chi$ (green dots) fails to capture the topology in the point-gapped phase, deviating from the Bloch invariant $P$ (red line). (b) Restoration of the bulk-boundary correspondence: By employing the quasi-reciprocal Hamiltonian $\tilde{H}$, the entanglement polarization $\chi(\tilde{H})$ (blue circles) is robustly quantized and matches the non-Bloch polarization $P_\beta$ (yellow triangles), confirming $P_\beta \equiv \chi(\tilde{H})\pmod 1$. (c) The GBZ (blue curve) at $t_1=0.8$. Its deviation from the unit circle (grey dashed line) induces long-range power-law hoppings in $\tilde{H}$. (d) Spatial localization of the entanglement edge modes ($\xi=0.5$). Parameters are fixed at $t_2=1, \gamma=0.4, t_3=0.2$, with system size $L=800$.}
\label{Fig1}
\end{figure}

We first demonstrate the robustness of our theory in the challenging point-gapped phase, where the strong NHSE drives the GBZ contour $C_\beta$ far from the unit circle, generating the long-range couplings discussed above. Figure~\ref{Fig1}(a) shows the EOS computed from the PBC ground state of the original Hamiltonian $H$ over a parameter sweep, $t_1 \in [0, 2.5]$. In the point-gapped region $0.7 < t_1 < 1.6$, the EOS $\xi_\mu$ fails to exhibit the characteristic $\xi=0.5$ modes, highlighting the breakdown of conventional Bloch theory in this regime~\cite{OrtegaTaberner2022, Loic2019}. Conversely, in the line-gapped regions ($0 < t_1 < 0.7$ topologically non-trivial, $1.6 < t_1 < 2.5$ trivial), the EOS accurately reflects the expected Bloch polarization values, $P = 0.5$ and $P = 0$, respectively.

In contrast, Fig.~\ref{Fig1}(b) presents the definitive validation of our theoretical prediction. For each $t_1$, we compute the non-circular GBZ contour $C_\beta$ (Fig.~\ref{Fig1}(c) for $t_1 = 0.8$), construct the quasi-reciprocal Hamiltonian $\tilde{H}$, and evaluate $P_\beta$ and $\chi(\tilde{H})$ (see Supplementary Material Section~\textcolor{blue}{\hyperref[sec: S4]{S4}} for the robust projection method used to handle spectral degeneracies). For each $t_1$, we compute the non-circular GBZ contour $C_\beta$ (Fig.~\ref{Fig1}(c) for $t_1 = 0.8$), construct the quasi-reciprocal Hamiltonian $\tilde{H}$, and evaluate $P_\beta$ and $\chi(\tilde{H})$. The entanglement polarization $\chi(\tilde{H})$ robustly diagnoses the topological phase: it clearly distinguishes between trivial ($\chi(\tilde{H}) = 0$) and non-trivial ($\chi(\tilde{H}) = 0.5$) phases, with a sharp transition at $t_1 = 1.33$, where both $\chi(\tilde{H})$ and $P_\beta$ jump discontinuously. Crucially, this equivalence holds precisely in the regime where $\tilde{H}$ becomes non-local. While the divergence of the position variance in this regime would render the Resta polarization ill-defined, $\chi(\tilde{H})$ remains strictly quantized, confirming its immunity to the breakdown of locality. The topological character of the non-zero $\chi(\tilde{H})$ plateau is confirmed by the spatial localization of the two $\xi = 0.5$ entanglement edge states (Fig.~\ref{Fig1}(d)), which are exponentially localized at the system boundaries.

These results provide compelling verification of the fundamental equivalence between the non-Bloch topological invariant $P_\beta$ and its real-space entanglement counterpart $\chi(\tilde{H})$. As demonstrated in Fig.~\ref{Fig1}, this equivalence, $P_\beta(H) \equiv \chi(\tilde{H})\pmod 1$, remains valid even under the non-local conditions induced by a non-circular GBZ. This success confirms that the topological protection of $\chi(\tilde{H})$ is governed by the Fredholm index of the Toeplitz correlation matrix. As detailed in Supplementary Material Section~\textcolor{blue}{\hyperref[sec: S3]{S3}}, this index remains robust against algebraic decay ($\alpha > 1$) of the hoppings, ensuring topological quantization even when exponential locality is lost. Notably, the chiral symmetry-imposed $\xi \leftrightarrow 1-\xi$ pairing in the EOS remains robust for all modes, restoring the EOS-topology correspondence that is otherwise broken in the point-gapped phase (see Supplementary Material Section~\textcolor{blue}{\hyperref[sec: S6]{S6}} for details).

\begin{figure}[t]
\centering
\includegraphics[width=0.4\textwidth,keepaspectratio]{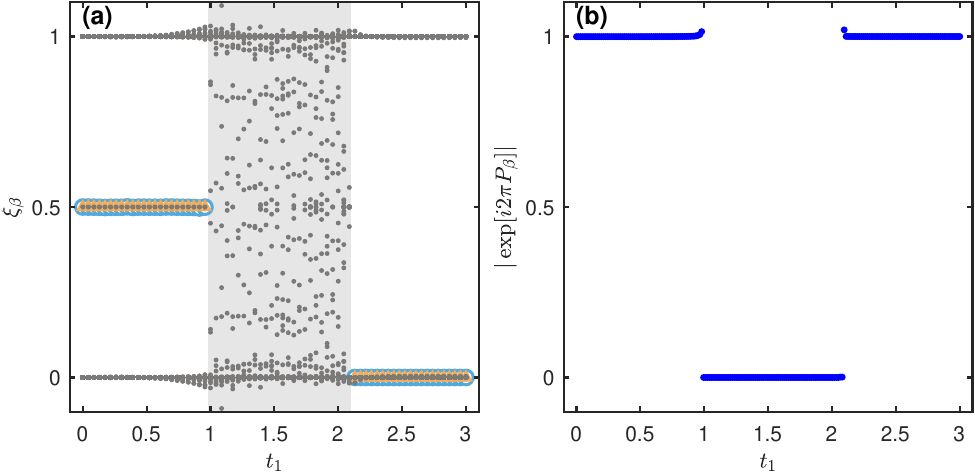}
\caption{Topological phase transition and correspondence breakdown. (a) Verification of the correspondence: The coincidence of the non-Bloch polarization $P_\beta$ (yellow triangles) and the real-space entanglement polarization $\chi(\tilde{H})$ (blue circles) validates the identity $P_\beta \equiv \chi(\tilde{H})\pmod 1$ in the gapped regions. The grey dots represent the entanglement orbital spectrum (EOS) of $\tilde{H}$. The shaded region indicates the gapless Topological Semimetal (TSM) phase. (b) Collapse of the invariant: The sharp drop of the non-Bloch Wilson loop norm $|W_\beta|$ to zero marks the phase transition, driven by Exceptional Points (EPs) intersecting the Generalized Brillouin Zone (GBZ) contour. Parameters are fixed at $t_2 = 1$, $\gamma = 1$, $t_3 = 0.3$, with system size $L=800$.}
\label{Fig2}
\end{figure}

A complete topological description must also capture the phase transition to a gapless topological semimetal (TSM) phase. Figure~\ref{Fig2} illustrates this transition for a different parameter set. In the gapped topological phase ($0 < t_1 < 0.95$), $\chi(\tilde{H})$ exhibits a quantized plateau protected by a finite entanglement gap. Within the TSM region (grey shaded), the OBC bulk gap closes, leading to a simultaneous closure of the entanglement gap of $\tilde{H}$. As a result, the entanglement spectrum $\xi$ delocalizes across $[0, 1]$, rendering $\chi(\tilde{H})$ ill-defined, a behavior directly analogous to the bulk gap closing in Hermitian transitions.

Crucially, this breakdown coincides exactly with the failure of the non-Bloch polarization $P_\beta$ in the TSM region. Figure~\ref{Fig2}(b) shows that $|W_\beta|$ remains pinned to unity in the gapped phase but collapses abruptly to zero in the TSM region. This collapse arises because EPs lie directly on the GBZ integration path~\cite{Yokomizo2019, Yokomizo2020}. At these EPs, the biorthogonal overlap $\langle u^L(\beta) \vert u^R(\beta) \rangle$ vanishes. Since the Wilson loop $W_\beta = \vert W_\beta\vert e^{i 2\pi P_\beta}$ is defined as the product of these overlaps along the GBZ, a single vanishing factor forces $\vert W_\beta\vert \to 0$, rendering the phase $P_\beta$ mathematically indeterminate. The simultaneous collapse of the geometric invariant $P_\beta$ (due to EPs) and the algebraic invariant $\chi$ (due to gap closing) reinforces their fundamental duality. While $P_\beta$ fails due to the geometric singularity of the GBZ, $\chi(\tilde{H})$ correctly diagnoses the phase transition through the closure of the spectral gap, offering a more physically transparent picture.

{\it Discussions and Outlooks.}---
The established equivalence $P_\beta \equiv \chi(\tilde{H}) \pmod 1$ holds profound significance for non-Hermitian physics, fundamentally unifying its two core paradigms. It rigorously connects the abstract momentum-space geometric phase $P_\beta$, calculated on the GBZ, with a concrete real-space entanglement invariant $\chi(\tilde{H})$ of the quasi-reciprocal Hamiltonian. This correspondence provides a direct real-space physical picture for the non-Bloch topological invariant, encoded explicitly in the entanglement structure of $\tilde{H}$. Crucially, the true power of this framework is revealed in the non-local regime induced by a non-circular GBZ. Here, the divergence of position variance violates the nearsightedness principle, rendering the geometric Resta polarization ill-defined. In stark contrast, $\chi(\tilde{H})$ remains robustly quantized. As analytically derived in Supplementary Material Section~\textcolor{blue}{\hyperref[sec: S3]{S3}}, this robustness stems from the Fredholm index of the underlying Toeplitz structure, which protects $\chi(\tilde{H})$ even when exponential locality is lost. This demonstrates that $P_\beta \equiv \chi(\tilde H)\pmod 1$ represents a topological identity more fundamental than the constraint of locality, establishing entanglement as the uniquely robust real-space diagnostic for non-Hermitian topology.

Conceptually, this framework provides a complete and physically intuitive picture of topological phase transitions. As demonstrated in the TSM regime, the simultaneous breakdown of both invariants is not a deficiency, but rather compelling evidence of their fundamental equivalence. Unlike the abstract collapse of the non-Bloch Wilson loop, the closure of the entanglement gap offers a tangible diagnostic, restoring the standard intuition of "gap-closing" topology to non-Hermitian systems. Thus, $\chi(\tilde{H})$ serves not only as a robust invariant for gapped phases but also as a transparent probe for critical transitions.

Our framework transforms the study of non-Bloch topology, moving from abstract GBZ calculations to the measurement of concrete physical observables. This represents a crucial conceptual leap: while $P_\beta$ is confined to a complex manifold, $\chi(\tilde{H})$ is encoded in the tangible many-body entanglement of the system. Experimentally, topoelectrical circuits~\cite{Ezawa2019, Chen2024a, Sahin2025, Yuan2023} are particularly well-suited for this task, as the non-local couplings required for $\tilde{H}$ can be realized through hard-wired connections, permitting the extraction of $\chi(\tilde{H})$ from admittance data. Furthermore, superconducting qubit arrays~\cite{Ren2025} allow for the digital simulation of non-local Hamiltonians, where state tomography can recover the entanglement spectrum. Additionally, synthetic dimensions in photonic~\cite{Parto2023,Gao2024} and atomic systems~\cite{Liang2022, Guo2022} offer versatile routes to achieve the necessary long-range connectivity. Measuring $\chi(\tilde{H})$ across these platforms would provide definitive experimental proof of non-Bloch topology in the regime where locality breaks down.

In summary, our work demonstrates that the entanglement polarization $\chi(\tilde H)$, derived from the effective entanglement Hamiltonian, acts as a direct and robust real-space bulk probe for the non-Bloch topological invariant $P_\beta$. Its theoretically protected robustness against non-locality stands in stark contrast to the fragility of geometric invariants, effectively bridging momentum-space non-Bloch band theory with real-space topological characterization. By establishing the equivalence $P_\beta \equiv \chi(\tilde H)\pmod 1$, we complete the generalized bulk-boundary correspondence for non-Hermitian systems, identifying entanglement as the essential key to unlocking topology beyond the limits of locality. This framework provides a solid foundation for future studies of complex scenarios, including interacting and higher-dimensional non-Hermitian topological matter, where entanglement-based approaches are particularly critical.

{\it Acknowledgements.}---This work is supported by the National Natural Science Foundation of China (Grant No. 11975175).

\bibliography{references_main.bib}

\begin{thebibliography}{91}%
\makeatletter
\providecommand \@ifxundefined [1]{%
 \@ifx{#1\undefined}
}%
\providecommand \@ifnum [1]{%
 \ifnum #1\expandafter \@firstoftwo
 \else \expandafter \@secondoftwo
 \fi
}%
\providecommand \@ifx [1]{%
 \ifx #1\expandafter \@firstoftwo
 \else \expandafter \@secondoftwo
 \fi
}%
\providecommand \natexlab [1]{#1}%
\providecommand \enquote  [1]{``#1''}%
\providecommand \bibnamefont  [1]{#1}%
\providecommand \bibfnamefont [1]{#1}%
\providecommand \citenamefont [1]{#1}%
\providecommand \href@noop [0]{\@secondoftwo}%
\providecommand \href [0]{\begingroup \@sanitize@url \@href}%
\providecommand \@href[1]{\@@startlink{#1}\@@href}%
\providecommand \@@href[1]{\endgroup#1\@@endlink}%
\providecommand \@sanitize@url [0]{\catcode `\\12\catcode `\$12\catcode `\&12\catcode `\#12\catcode `\^12\catcode `\_12\catcode `\%12\relax}%
\providecommand \@@startlink[1]{}%
\providecommand \@@endlink[0]{}%
\providecommand \url  [0]{\begingroup\@sanitize@url \@url }%
\providecommand \@url [1]{\endgroup\@href {#1}{\urlprefix }}%
\providecommand \urlprefix  [0]{URL }%
\providecommand \Eprint [0]{\href }%
\providecommand \doibase [0]{https://doi.org/}%
\providecommand \selectlanguage [0]{\@gobble}%
\providecommand \bibinfo  [0]{\@secondoftwo}%
\providecommand \bibfield  [0]{\@secondoftwo}%
\providecommand \translation [1]{[#1]}%
\providecommand \BibitemOpen [0]{}%
\providecommand \bibitemStop [0]{}%
\providecommand \bibitemNoStop [0]{.\EOS\space}%
\providecommand \EOS [0]{\spacefactor3000\relax}%
\providecommand \BibitemShut  [1]{\csname bibitem#1\endcsname}%
\let\auto@bib@innerbib\@empty
\bibitem [{\citenamefont {Bender}\ and\ \citenamefont {Boettcher}(1998)}]{Bender1998}%
  \BibitemOpen
  \bibfield  {author} {\bibinfo {author} {\bibfnamefont {C.~M.}\ \bibnamefont {Bender}}\ and\ \bibinfo {author} {\bibfnamefont {S.}~\bibnamefont {Boettcher}},\ }\bibfield  {title} {\bibinfo {title} {Real spectra in non-{H}ermitian hamiltonians having $\mathcal{P}\mathcal{T}$ symmetry},\ }\href {https://doi.org/10.1103/PhysRevLett.80.5243} {\bibfield  {journal} {\bibinfo  {journal} {Phys. Rev. Lett.}\ }\textbf {\bibinfo {volume} {80}},\ \bibinfo {pages} {5243} (\bibinfo {year} {1998})}\BibitemShut {NoStop}%
\bibitem [{\citenamefont {Ashida}\ \emph {et~al.}(2020)\citenamefont {Ashida}, \citenamefont {Gong},\ and\ \citenamefont {Ueda}}]{Ashida2020}%
  \BibitemOpen
  \bibfield  {author} {\bibinfo {author} {\bibfnamefont {Y.}~\bibnamefont {Ashida}}, \bibinfo {author} {\bibfnamefont {Z.}~\bibnamefont {Gong}},\ and\ \bibinfo {author} {\bibfnamefont {M.}~\bibnamefont {Ueda}},\ }\bibfield  {title} {\bibinfo {title} {Non-{H}ermitian physics},\ }\href {https://doi.org/10.1080/00018732.2021.1876991} {\bibfield  {journal} {\bibinfo  {journal} {Adv. Phys.}\ }\textbf {\bibinfo {volume} {69}},\ \bibinfo {pages} {249} (\bibinfo {year} {2020})}\BibitemShut {NoStop}%
\bibitem [{\citenamefont {El-Ganainy}\ \emph {et~al.}(2018)\citenamefont {El-Ganainy}, \citenamefont {Makris}, \citenamefont {Khajavikhan}, \citenamefont {Musslimani}, \citenamefont {Rotter},\ and\ \citenamefont {Christodoulides}}]{ElGanainy2018}%
  \BibitemOpen
  \bibfield  {author} {\bibinfo {author} {\bibfnamefont {R.}~\bibnamefont {El-Ganainy}}, \bibinfo {author} {\bibfnamefont {K.~G.}\ \bibnamefont {Makris}}, \bibinfo {author} {\bibfnamefont {M.}~\bibnamefont {Khajavikhan}}, \bibinfo {author} {\bibfnamefont {Z.~H.}\ \bibnamefont {Musslimani}}, \bibinfo {author} {\bibfnamefont {S.}~\bibnamefont {Rotter}},\ and\ \bibinfo {author} {\bibfnamefont {D.~N.}\ \bibnamefont {Christodoulides}},\ }\bibfield  {title} {\bibinfo {title} {Non-{H}ermitian physics and {PT} symmetry},\ }\href {https://doi.org/10.1038/nphys4323} {\bibfield  {journal} {\bibinfo  {journal} {Nat. Phys.}\ }\textbf {\bibinfo {volume} {14}},\ \bibinfo {pages} {11} (\bibinfo {year} {2018})}\BibitemShut {NoStop}%
\bibitem [{\citenamefont {Gong}\ \emph {et~al.}(2018)\citenamefont {Gong}, \citenamefont {Ashida}, \citenamefont {Kawabata}, \citenamefont {Takasan}, \citenamefont {Higashikawa},\ and\ \citenamefont {Ueda}}]{Gong2018}%
  \BibitemOpen
  \bibfield  {author} {\bibinfo {author} {\bibfnamefont {Z.}~\bibnamefont {Gong}}, \bibinfo {author} {\bibfnamefont {Y.}~\bibnamefont {Ashida}}, \bibinfo {author} {\bibfnamefont {K.}~\bibnamefont {Kawabata}}, \bibinfo {author} {\bibfnamefont {K.}~\bibnamefont {Takasan}}, \bibinfo {author} {\bibfnamefont {S.}~\bibnamefont {Higashikawa}},\ and\ \bibinfo {author} {\bibfnamefont {M.}~\bibnamefont {Ueda}},\ }\bibfield  {title} {\bibinfo {title} {Topological phases of non-{H}ermitian systems},\ }\href {https://doi.org/10.1103/PhysRevX.8.031079} {\bibfield  {journal} {\bibinfo  {journal} {Phys. Rev. X}\ }\textbf {\bibinfo {volume} {8}},\ \bibinfo {pages} {031079} (\bibinfo {year} {2018})}\BibitemShut {NoStop}%
\bibitem [{\citenamefont {Bergholtz}\ \emph {et~al.}(2021)\citenamefont {Bergholtz}, \citenamefont {Budich},\ and\ \citenamefont {Kunst}}]{Bergholtz2021}%
  \BibitemOpen
  \bibfield  {author} {\bibinfo {author} {\bibfnamefont {E.~J.}\ \bibnamefont {Bergholtz}}, \bibinfo {author} {\bibfnamefont {J.~C.}\ \bibnamefont {Budich}},\ and\ \bibinfo {author} {\bibfnamefont {F.~K.}\ \bibnamefont {Kunst}},\ }\bibfield  {title} {\bibinfo {title} {Exceptional topology of non-{H}ermitian systems},\ }\href {https://doi.org/10.1103/RevModPhys.93.015005} {\bibfield  {journal} {\bibinfo  {journal} {Rev. Mod. Phys.}\ }\textbf {\bibinfo {volume} {93}},\ \bibinfo {pages} {015005} (\bibinfo {year} {2021})}\BibitemShut {NoStop}%
\bibitem [{\citenamefont {Okuma}\ and\ \citenamefont {Sato}(2023)}]{Okuma2023}%
  \BibitemOpen
  \bibfield  {author} {\bibinfo {author} {\bibfnamefont {N.}~\bibnamefont {Okuma}}\ and\ \bibinfo {author} {\bibfnamefont {M.}~\bibnamefont {Sato}},\ }\bibfield  {title} {\bibinfo {title} {Non-{H}ermitian topological phenomena: A review},\ }\href {https://doi.org/https://doi.org/10.1146/annurev-conmatphys-040521-033133} {\bibfield  {journal} {\bibinfo  {journal} {Annu. Rev. Condens. Matter Phys.}\ }\textbf {\bibinfo {volume} {14}},\ \bibinfo {pages} {83} (\bibinfo {year} {2023})}\BibitemShut {NoStop}%
\bibitem [{\citenamefont {Wang}\ \emph {et~al.}(2021)\citenamefont {Wang}, \citenamefont {Zhang}, \citenamefont {Hua}, \citenamefont {Lei}, \citenamefont {Lu},\ and\ \citenamefont {Chen}}]{Wang2021}%
  \BibitemOpen
  \bibfield  {author} {\bibinfo {author} {\bibfnamefont {H.}~\bibnamefont {Wang}}, \bibinfo {author} {\bibfnamefont {X.}~\bibnamefont {Zhang}}, \bibinfo {author} {\bibfnamefont {J.}~\bibnamefont {Hua}}, \bibinfo {author} {\bibfnamefont {D.}~\bibnamefont {Lei}}, \bibinfo {author} {\bibfnamefont {M.}~\bibnamefont {Lu}},\ and\ \bibinfo {author} {\bibfnamefont {Y.}~\bibnamefont {Chen}},\ }\bibfield  {title} {\bibinfo {title} {Topological physics of non-{H}ermitian optics and photonics: a review},\ }\href {https://doi.org/10.1088/2040-8986/ac2e15} {\bibfield  {journal} {\bibinfo  {journal} {J. Opt.}\ }\textbf {\bibinfo {volume} {23}},\ \bibinfo {pages} {123001} (\bibinfo {year} {2021})}\BibitemShut {NoStop}%
\bibitem [{\citenamefont {Bender}(2007)}]{Bender2007}%
  \BibitemOpen
  \bibfield  {author} {\bibinfo {author} {\bibfnamefont {C.~M.}\ \bibnamefont {Bender}},\ }\bibfield  {title} {\bibinfo {title} {Making sense of non-{H}ermitian {H}amiltonians},\ }\href {https://doi.org/10.1088/0034-4885/70/6/R03} {\bibfield  {journal} {\bibinfo  {journal} {Rep. Prog. Phys.}\ }\textbf {\bibinfo {volume} {70}},\ \bibinfo {pages} {947} (\bibinfo {year} {2007})}\BibitemShut {NoStop}%
\bibitem [{\citenamefont {Lu}\ \emph {et~al.}(2014)\citenamefont {Lu}, \citenamefont {Joannopoulos},\ and\ \citenamefont {Soljačić}}]{Lu2014}%
  \BibitemOpen
  \bibfield  {author} {\bibinfo {author} {\bibfnamefont {L.}~\bibnamefont {Lu}}, \bibinfo {author} {\bibfnamefont {J.~D.}\ \bibnamefont {Joannopoulos}},\ and\ \bibinfo {author} {\bibfnamefont {M.}~\bibnamefont {Soljačić}},\ }\bibfield  {title} {\bibinfo {title} {Topological photonics},\ }\href {https://doi.org/10.1038/nphoton.2014.248} {\bibfield  {journal} {\bibinfo  {journal} {Nat. Photon.}\ }\textbf {\bibinfo {volume} {8}},\ \bibinfo {pages} {821} (\bibinfo {year} {2014})}\BibitemShut {NoStop}%
\bibitem [{\citenamefont {Miri}\ and\ \citenamefont {Alù}(2019)}]{Miri2019}%
  \BibitemOpen
  \bibfield  {author} {\bibinfo {author} {\bibfnamefont {M.-A.}\ \bibnamefont {Miri}}\ and\ \bibinfo {author} {\bibfnamefont {A.}~\bibnamefont {Alù}},\ }\bibfield  {title} {\bibinfo {title} {Exceptional points in optics and photonics},\ }\href {https://doi.org/10.1126/science.aar7709} {\bibfield  {journal} {\bibinfo  {journal} {Science}\ }\textbf {\bibinfo {volume} {363}},\ \bibinfo {pages} {eaar7709} (\bibinfo {year} {2019})}\BibitemShut {NoStop}%
\bibitem [{\citenamefont {Heiss}(2012)}]{Heiss2012}%
  \BibitemOpen
  \bibfield  {author} {\bibinfo {author} {\bibfnamefont {W.~D.}\ \bibnamefont {Heiss}},\ }\bibfield  {title} {\bibinfo {title} {The physics of exceptional points},\ }\href {https://doi.org/10.1088/1751-8113/45/44/444016} {\bibfield  {journal} {\bibinfo  {journal} {J. Phys. A Math. Theor.}\ }\textbf {\bibinfo {volume} {45}},\ \bibinfo {pages} {444016} (\bibinfo {year} {2012})}\BibitemShut {NoStop}%
\bibitem [{\citenamefont {Can\'os~Valero}\ \emph {et~al.}(2025)\citenamefont {Can\'os~Valero}, \citenamefont {Sztranyovszky}, \citenamefont {Muljarov}, \citenamefont {Bogdanov},\ and\ \citenamefont {Weiss}}]{CanosValero2025}%
  \BibitemOpen
  \bibfield  {author} {\bibinfo {author} {\bibfnamefont {A.}~\bibnamefont {Can\'os~Valero}}, \bibinfo {author} {\bibfnamefont {Z.}~\bibnamefont {Sztranyovszky}}, \bibinfo {author} {\bibfnamefont {E.~A.}\ \bibnamefont {Muljarov}}, \bibinfo {author} {\bibfnamefont {A.}~\bibnamefont {Bogdanov}},\ and\ \bibinfo {author} {\bibfnamefont {T.}~\bibnamefont {Weiss}},\ }\bibfield  {title} {\bibinfo {title} {Exceptional bound states in the continuum},\ }\href {https://doi.org/10.1103/PhysRevLett.134.103802} {\bibfield  {journal} {\bibinfo  {journal} {Phys. Rev. Lett.}\ }\textbf {\bibinfo {volume} {134}},\ \bibinfo {pages} {103802} (\bibinfo {year} {2025})}\BibitemShut {NoStop}%
\bibitem [{\citenamefont {Yokomizo}\ and\ \citenamefont {Murakami}(2020)}]{Yokomizo2020}%
  \BibitemOpen
  \bibfield  {author} {\bibinfo {author} {\bibfnamefont {K.}~\bibnamefont {Yokomizo}}\ and\ \bibinfo {author} {\bibfnamefont {S.}~\bibnamefont {Murakami}},\ }\bibfield  {title} {\bibinfo {title} {Topological semimetal phase with exceptional points in one-dimensional non-{H}ermitian systems},\ }\href {https://doi.org/10.1103/PhysRevResearch.2.043045} {\bibfield  {journal} {\bibinfo  {journal} {Phys. Rev. Res.}\ }\textbf {\bibinfo {volume} {2}},\ \bibinfo {pages} {043045} (\bibinfo {year} {2020})}\BibitemShut {NoStop}%
\bibitem [{\citenamefont {Zhen}\ \emph {et~al.}(2015)\citenamefont {Zhen}, \citenamefont {Hsu}, \citenamefont {Igarashi}, \citenamefont {Lu}, \citenamefont {Kaminer}, \citenamefont {Pick}, \citenamefont {Chua}, \citenamefont {Joannopoulos},\ and\ \citenamefont {Soljačić}}]{Zhen2015}%
  \BibitemOpen
  \bibfield  {author} {\bibinfo {author} {\bibfnamefont {B.}~\bibnamefont {Zhen}}, \bibinfo {author} {\bibfnamefont {C.~W.}\ \bibnamefont {Hsu}}, \bibinfo {author} {\bibfnamefont {Y.}~\bibnamefont {Igarashi}}, \bibinfo {author} {\bibfnamefont {L.}~\bibnamefont {Lu}}, \bibinfo {author} {\bibfnamefont {I.}~\bibnamefont {Kaminer}}, \bibinfo {author} {\bibfnamefont {A.}~\bibnamefont {Pick}}, \bibinfo {author} {\bibfnamefont {S.-L.}\ \bibnamefont {Chua}}, \bibinfo {author} {\bibfnamefont {J.~D.}\ \bibnamefont {Joannopoulos}},\ and\ \bibinfo {author} {\bibfnamefont {M.}~\bibnamefont {Soljačić}},\ }\bibfield  {title} {\bibinfo {title} {Spawning rings of exceptional points out of dirac cones},\ }\href {https://doi.org/10.1038/nature14889} {\bibfield  {journal} {\bibinfo  {journal} {Nature}\ }\textbf {\bibinfo {volume} {525}},\ \bibinfo {pages} {354} (\bibinfo {year} {2015})}\BibitemShut {NoStop}%
\bibitem [{\citenamefont {Doppler}\ \emph {et~al.}(2016)\citenamefont {Doppler}, \citenamefont {Mailybaev}, \citenamefont {Böhm}, \citenamefont {Kuhl}, \citenamefont {Girschik}, \citenamefont {Libisch}, \citenamefont {Milburn}, \citenamefont {Rabl}, \citenamefont {Moiseyev},\ and\ \citenamefont {Rotter}}]{Doppler2016}%
  \BibitemOpen
  \bibfield  {author} {\bibinfo {author} {\bibfnamefont {J.}~\bibnamefont {Doppler}}, \bibinfo {author} {\bibfnamefont {A.~A.}\ \bibnamefont {Mailybaev}}, \bibinfo {author} {\bibfnamefont {J.}~\bibnamefont {Böhm}}, \bibinfo {author} {\bibfnamefont {U.}~\bibnamefont {Kuhl}}, \bibinfo {author} {\bibfnamefont {A.}~\bibnamefont {Girschik}}, \bibinfo {author} {\bibfnamefont {F.}~\bibnamefont {Libisch}}, \bibinfo {author} {\bibfnamefont {T.~J.}\ \bibnamefont {Milburn}}, \bibinfo {author} {\bibfnamefont {P.}~\bibnamefont {Rabl}}, \bibinfo {author} {\bibfnamefont {N.}~\bibnamefont {Moiseyev}},\ and\ \bibinfo {author} {\bibfnamefont {S.}~\bibnamefont {Rotter}},\ }\bibfield  {title} {\bibinfo {title} {Dynamically encircling an exceptional point for asymmetric mode switching},\ }\href {https://doi.org/10.1038/nature18605} {\bibfield  {journal} {\bibinfo  {journal} {Nature}\ }\textbf {\bibinfo {volume} {537}},\ \bibinfo {pages} {76} (\bibinfo {year} {2016})}\BibitemShut {NoStop}%
\bibitem [{\citenamefont {Budich}\ \emph {et~al.}(2019)\citenamefont {Budich}, \citenamefont {Carlstr\"om}, \citenamefont {Kunst},\ and\ \citenamefont {Bergholtz}}]{Budich2019}%
  \BibitemOpen
  \bibfield  {author} {\bibinfo {author} {\bibfnamefont {J.~C.}\ \bibnamefont {Budich}}, \bibinfo {author} {\bibfnamefont {J.}~\bibnamefont {Carlstr\"om}}, \bibinfo {author} {\bibfnamefont {F.~K.}\ \bibnamefont {Kunst}},\ and\ \bibinfo {author} {\bibfnamefont {E.~J.}\ \bibnamefont {Bergholtz}},\ }\bibfield  {title} {\bibinfo {title} {Symmetry-protected nodal phases in non-{H}ermitian systems},\ }\href {https://doi.org/10.1103/PhysRevB.99.041406} {\bibfield  {journal} {\bibinfo  {journal} {Phys. Rev. B}\ }\textbf {\bibinfo {volume} {99}},\ \bibinfo {pages} {041406} (\bibinfo {year} {2019})}\BibitemShut {NoStop}%
\bibitem [{\citenamefont {Kawabata}\ \emph {et~al.}(2021)\citenamefont {Kawabata}, \citenamefont {Shiozaki},\ and\ \citenamefont {Ryu}}]{Kawabata2021}%
  \BibitemOpen
  \bibfield  {author} {\bibinfo {author} {\bibfnamefont {K.}~\bibnamefont {Kawabata}}, \bibinfo {author} {\bibfnamefont {K.}~\bibnamefont {Shiozaki}},\ and\ \bibinfo {author} {\bibfnamefont {S.}~\bibnamefont {Ryu}},\ }\bibfield  {title} {\bibinfo {title} {Topological field theory of non-hermitian systems},\ }\href {https://doi.org/10.1103/PhysRevLett.126.216405} {\bibfield  {journal} {\bibinfo  {journal} {Phys. Rev. Lett.}\ }\textbf {\bibinfo {volume} {126}},\ \bibinfo {pages} {216405} (\bibinfo {year} {2021})}\BibitemShut {NoStop}%
\bibitem [{\citenamefont {Xiong}\ \emph {et~al.}(2021)\citenamefont {Xiong}, \citenamefont {Li}, \citenamefont {Song}, \citenamefont {Chen}, \citenamefont {Zhang},\ and\ \citenamefont {Wang}}]{Xiong2021}%
  \BibitemOpen
  \bibfield  {author} {\bibinfo {author} {\bibfnamefont {W.}~\bibnamefont {Xiong}}, \bibinfo {author} {\bibfnamefont {Z.}~\bibnamefont {Li}}, \bibinfo {author} {\bibfnamefont {Y.}~\bibnamefont {Song}}, \bibinfo {author} {\bibfnamefont {J.}~\bibnamefont {Chen}}, \bibinfo {author} {\bibfnamefont {G.-Q.}\ \bibnamefont {Zhang}},\ and\ \bibinfo {author} {\bibfnamefont {M.}~\bibnamefont {Wang}},\ }\bibfield  {title} {\bibinfo {title} {Higher-order exceptional point in a pseudo-hermitian cavity optomechanical system},\ }\href {https://doi.org/10.1103/PhysRevA.104.063508} {\bibfield  {journal} {\bibinfo  {journal} {Phys. Rev. A}\ }\textbf {\bibinfo {volume} {104}},\ \bibinfo {pages} {063508} (\bibinfo {year} {2021})}\BibitemShut {NoStop}%
\bibitem [{\citenamefont {Luitz}\ and\ \citenamefont {Piazza}(2019)}]{Luitz2019}%
  \BibitemOpen
  \bibfield  {author} {\bibinfo {author} {\bibfnamefont {D.~J.}\ \bibnamefont {Luitz}}\ and\ \bibinfo {author} {\bibfnamefont {F.}~\bibnamefont {Piazza}},\ }\bibfield  {title} {\bibinfo {title} {Exceptional points and the topology of quantum many-body spectra},\ }\href {https://doi.org/10.1103/PhysRevResearch.1.033051} {\bibfield  {journal} {\bibinfo  {journal} {Phys. Rev. Res.}\ }\textbf {\bibinfo {volume} {1}},\ \bibinfo {pages} {033051} (\bibinfo {year} {2019})}\BibitemShut {NoStop}%
\bibitem [{\citenamefont {Cai}\ \emph {et~al.}(2022)\citenamefont {Cai}, \citenamefont {Jin}, \citenamefont {Li}, \citenamefont {Rabczuk}, \citenamefont {Pennec}, \citenamefont {Djafari-Rouhani},\ and\ \citenamefont {Zhuang}}]{Cai2022}%
  \BibitemOpen
  \bibfield  {author} {\bibinfo {author} {\bibfnamefont {R.}~\bibnamefont {Cai}}, \bibinfo {author} {\bibfnamefont {Y.}~\bibnamefont {Jin}}, \bibinfo {author} {\bibfnamefont {Y.}~\bibnamefont {Li}}, \bibinfo {author} {\bibfnamefont {T.}~\bibnamefont {Rabczuk}}, \bibinfo {author} {\bibfnamefont {Y.}~\bibnamefont {Pennec}}, \bibinfo {author} {\bibfnamefont {B.}~\bibnamefont {Djafari-Rouhani}},\ and\ \bibinfo {author} {\bibfnamefont {X.}~\bibnamefont {Zhuang}},\ }\bibfield  {title} {\bibinfo {title} {Exceptional points and skin modes in non-hermitian metabeams},\ }\href {https://doi.org/10.1103/PhysRevApplied.18.014067} {\bibfield  {journal} {\bibinfo  {journal} {Phys. Rev. Appl.}\ }\textbf {\bibinfo {volume} {18}},\ \bibinfo {pages} {014067} (\bibinfo {year} {2022})}\BibitemShut {NoStop}%
\bibitem [{\citenamefont {Hu}\ \emph {et~al.}(2017)\citenamefont {Hu}, \citenamefont {Wang}, \citenamefont {Shum},\ and\ \citenamefont {Chong}}]{Hu2017}%
  \BibitemOpen
  \bibfield  {author} {\bibinfo {author} {\bibfnamefont {W.}~\bibnamefont {Hu}}, \bibinfo {author} {\bibfnamefont {H.}~\bibnamefont {Wang}}, \bibinfo {author} {\bibfnamefont {P.~P.}\ \bibnamefont {Shum}},\ and\ \bibinfo {author} {\bibfnamefont {Y.~D.}\ \bibnamefont {Chong}},\ }\bibfield  {title} {\bibinfo {title} {Exceptional points in a non-hermitian topological pump},\ }\href {https://doi.org/10.1103/PhysRevB.95.184306} {\bibfield  {journal} {\bibinfo  {journal} {Phys. Rev. B}\ }\textbf {\bibinfo {volume} {95}},\ \bibinfo {pages} {184306} (\bibinfo {year} {2017})}\BibitemShut {NoStop}%
\bibitem [{\citenamefont {Yang}\ \emph {et~al.}(2021)\citenamefont {Yang}, \citenamefont {Schnyder}, \citenamefont {Hu},\ and\ \citenamefont {Chiu}}]{Yang2021}%
  \BibitemOpen
  \bibfield  {author} {\bibinfo {author} {\bibfnamefont {Z.}~\bibnamefont {Yang}}, \bibinfo {author} {\bibfnamefont {A.~P.}\ \bibnamefont {Schnyder}}, \bibinfo {author} {\bibfnamefont {J.}~\bibnamefont {Hu}},\ and\ \bibinfo {author} {\bibfnamefont {C.-K.}\ \bibnamefont {Chiu}},\ }\bibfield  {title} {\bibinfo {title} {Fermion doubling theorems in two-dimensional non-hermitian systems for fermi points and exceptional points},\ }\href {https://doi.org/10.1103/PhysRevLett.126.086401} {\bibfield  {journal} {\bibinfo  {journal} {Phys. Rev. Lett.}\ }\textbf {\bibinfo {volume} {126}},\ \bibinfo {pages} {086401} (\bibinfo {year} {2021})}\BibitemShut {NoStop}%
\bibitem [{\citenamefont {Mandal}\ and\ \citenamefont {Bergholtz}(2021)}]{Mandal2021}%
  \BibitemOpen
  \bibfield  {author} {\bibinfo {author} {\bibfnamefont {I.}~\bibnamefont {Mandal}}\ and\ \bibinfo {author} {\bibfnamefont {E.~J.}\ \bibnamefont {Bergholtz}},\ }\bibfield  {title} {\bibinfo {title} {Symmetry and higher-order exceptional points},\ }\href {https://doi.org/10.1103/PhysRevLett.127.186601} {\bibfield  {journal} {\bibinfo  {journal} {Phys. Rev. Lett.}\ }\textbf {\bibinfo {volume} {127}},\ \bibinfo {pages} {186601} (\bibinfo {year} {2021})}\BibitemShut {NoStop}%
\bibitem [{\citenamefont {Xiao}\ \emph {et~al.}(2024)\citenamefont {Xiao}, \citenamefont {Chu}, \citenamefont {Lin}, \citenamefont {Lin}, \citenamefont {Yi}, \citenamefont {Cai},\ and\ \citenamefont {Xue}}]{Xiao2024a}%
  \BibitemOpen
  \bibfield  {author} {\bibinfo {author} {\bibfnamefont {L.}~\bibnamefont {Xiao}}, \bibinfo {author} {\bibfnamefont {Y.}~\bibnamefont {Chu}}, \bibinfo {author} {\bibfnamefont {Q.}~\bibnamefont {Lin}}, \bibinfo {author} {\bibfnamefont {H.}~\bibnamefont {Lin}}, \bibinfo {author} {\bibfnamefont {W.}~\bibnamefont {Yi}}, \bibinfo {author} {\bibfnamefont {J.}~\bibnamefont {Cai}},\ and\ \bibinfo {author} {\bibfnamefont {P.}~\bibnamefont {Xue}},\ }\bibfield  {title} {\bibinfo {title} {Non-hermitian sensing in the absence of exceptional points},\ }\href {https://doi.org/10.1103/PhysRevLett.133.180801} {\bibfield  {journal} {\bibinfo  {journal} {Phys. Rev. Lett.}\ }\textbf {\bibinfo {volume} {133}},\ \bibinfo {pages} {180801} (\bibinfo {year} {2024})}\BibitemShut {NoStop}%
\bibitem [{\citenamefont {Zhu}\ \emph {et~al.}(2018)\citenamefont {Zhu}, \citenamefont {Fang}, \citenamefont {Li}, \citenamefont {Sun}, \citenamefont {Li}, \citenamefont {Jing},\ and\ \citenamefont {Chen}}]{Zhu2018}%
  \BibitemOpen
  \bibfield  {author} {\bibinfo {author} {\bibfnamefont {W.}~\bibnamefont {Zhu}}, \bibinfo {author} {\bibfnamefont {X.}~\bibnamefont {Fang}}, \bibinfo {author} {\bibfnamefont {D.}~\bibnamefont {Li}}, \bibinfo {author} {\bibfnamefont {Y.}~\bibnamefont {Sun}}, \bibinfo {author} {\bibfnamefont {Y.}~\bibnamefont {Li}}, \bibinfo {author} {\bibfnamefont {Y.}~\bibnamefont {Jing}},\ and\ \bibinfo {author} {\bibfnamefont {H.}~\bibnamefont {Chen}},\ }\bibfield  {title} {\bibinfo {title} {Simultaneous observation of a topological edge state and exceptional point in an open and non-hermitian acoustic system},\ }\href {https://doi.org/10.1103/PhysRevLett.121.124501} {\bibfield  {journal} {\bibinfo  {journal} {Phys. Rev. Lett.}\ }\textbf {\bibinfo {volume} {121}},\ \bibinfo {pages} {124501} (\bibinfo {year} {2018})}\BibitemShut {NoStop}%
\bibitem [{\citenamefont {Yao}\ and\ \citenamefont {Wang}(2018)}]{Yao2018}%
  \BibitemOpen
  \bibfield  {author} {\bibinfo {author} {\bibfnamefont {S.}~\bibnamefont {Yao}}\ and\ \bibinfo {author} {\bibfnamefont {Z.}~\bibnamefont {Wang}},\ }\bibfield  {title} {\bibinfo {title} {Edge states and topological invariants of non-hermitian systems},\ }\href {https://doi.org/10.1103/PhysRevLett.121.086803} {\bibfield  {journal} {\bibinfo  {journal} {Phys. Rev. Lett.}\ }\textbf {\bibinfo {volume} {121}},\ \bibinfo {pages} {086803} (\bibinfo {year} {2018})}\BibitemShut {NoStop}%
\bibitem [{\citenamefont {Lee}\ and\ \citenamefont {Thomale}(2019)}]{Lee2019}%
  \BibitemOpen
  \bibfield  {author} {\bibinfo {author} {\bibfnamefont {C.~H.}\ \bibnamefont {Lee}}\ and\ \bibinfo {author} {\bibfnamefont {R.}~\bibnamefont {Thomale}},\ }\bibfield  {title} {\bibinfo {title} {Anatomy of skin modes and topology in non-{H}ermitian systems},\ }\href {https://doi.org/10.1103/PhysRevB.99.201103} {\bibfield  {journal} {\bibinfo  {journal} {Phys. Rev. B}\ }\textbf {\bibinfo {volume} {99}},\ \bibinfo {pages} {201103} (\bibinfo {year} {2019})}\BibitemShut {NoStop}%
\bibitem [{\citenamefont {Okuma}\ \emph {et~al.}(2020)\citenamefont {Okuma}, \citenamefont {Kawabata}, \citenamefont {Shiozaki},\ and\ \citenamefont {Sato}}]{Okuma2020}%
  \BibitemOpen
  \bibfield  {author} {\bibinfo {author} {\bibfnamefont {N.}~\bibnamefont {Okuma}}, \bibinfo {author} {\bibfnamefont {K.}~\bibnamefont {Kawabata}}, \bibinfo {author} {\bibfnamefont {K.}~\bibnamefont {Shiozaki}},\ and\ \bibinfo {author} {\bibfnamefont {M.}~\bibnamefont {Sato}},\ }\bibfield  {title} {\bibinfo {title} {Topological origin of non-hermitian skin effects},\ }\href {https://doi.org/10.1103/PhysRevLett.124.086801} {\bibfield  {journal} {\bibinfo  {journal} {Phys. Rev. Lett.}\ }\textbf {\bibinfo {volume} {124}},\ \bibinfo {pages} {086801} (\bibinfo {year} {2020})}\BibitemShut {NoStop}%
\bibitem [{\citenamefont {Lee}(2016)}]{Lee2016}%
  \BibitemOpen
  \bibfield  {author} {\bibinfo {author} {\bibfnamefont {T.~E.}\ \bibnamefont {Lee}},\ }\bibfield  {title} {\bibinfo {title} {Anomalous edge state in a non-hermitian lattice},\ }\href {https://doi.org/10.1103/PhysRevLett.116.133903} {\bibfield  {journal} {\bibinfo  {journal} {Phys. Rev. Lett.}\ }\textbf {\bibinfo {volume} {116}},\ \bibinfo {pages} {133903} (\bibinfo {year} {2016})}\BibitemShut {NoStop}%
\bibitem [{\citenamefont {Longhi}(2020{\natexlab{a}})}]{Longhi2020}%
  \BibitemOpen
  \bibfield  {author} {\bibinfo {author} {\bibfnamefont {S.}~\bibnamefont {Longhi}},\ }\bibfield  {title} {\bibinfo {title} {Unraveling the non-hermitian skin effect in dissipative systems},\ }\href {https://doi.org/10.1103/PhysRevB.102.201103} {\bibfield  {journal} {\bibinfo  {journal} {Phys. Rev. B}\ }\textbf {\bibinfo {volume} {102}},\ \bibinfo {pages} {201103} (\bibinfo {year} {2020}{\natexlab{a}})}\BibitemShut {NoStop}%
\bibitem [{\citenamefont {Zhang}\ \emph {et~al.}(2022)\citenamefont {Zhang}, \citenamefont {Yang},\ and\ \citenamefont {Fang}}]{Zhang2022a}%
  \BibitemOpen
  \bibfield  {author} {\bibinfo {author} {\bibfnamefont {K.}~\bibnamefont {Zhang}}, \bibinfo {author} {\bibfnamefont {Z.}~\bibnamefont {Yang}},\ and\ \bibinfo {author} {\bibfnamefont {C.}~\bibnamefont {Fang}},\ }\bibfield  {title} {\bibinfo {title} {Universal non-hermitian skin effect in two and higher dimensions},\ }\href {https://doi.org/10.1038/s41467-022-30161-6} {\bibfield  {journal} {\bibinfo  {journal} {Nat. Commun.}\ }\textbf {\bibinfo {volume} {13}},\ \bibinfo {pages} {2496} (\bibinfo {year} {2022})}\BibitemShut {NoStop}%
\bibitem [{\citenamefont {Zhang}\ \emph {et~al.}(2024)\citenamefont {Zhang}, \citenamefont {Yang},\ and\ \citenamefont {Sun}}]{Zhang2024a}%
  \BibitemOpen
  \bibfield  {author} {\bibinfo {author} {\bibfnamefont {K.}~\bibnamefont {Zhang}}, \bibinfo {author} {\bibfnamefont {Z.}~\bibnamefont {Yang}},\ and\ \bibinfo {author} {\bibfnamefont {K.}~\bibnamefont {Sun}},\ }\bibfield  {title} {\bibinfo {title} {Edge theory of non-hermitian skin modes in higher dimensions},\ }\href {https://doi.org/10.1103/PhysRevB.109.165127} {\bibfield  {journal} {\bibinfo  {journal} {Phys. Rev. B}\ }\textbf {\bibinfo {volume} {109}},\ \bibinfo {pages} {165127} (\bibinfo {year} {2024})}\BibitemShut {NoStop}%
\bibitem [{\citenamefont {Hou}\ \emph {et~al.}(2024)\citenamefont {Hou}, \citenamefont {Wu}, \citenamefont {Li}, \citenamefont {Basit}, \citenamefont {Wei}, \citenamefont {Chen}, \citenamefont {Grelu},\ and\ \citenamefont {Ni}}]{Hou2024}%
  \BibitemOpen
  \bibfield  {author} {\bibinfo {author} {\bibfnamefont {C.}~\bibnamefont {Hou}}, \bibinfo {author} {\bibfnamefont {G.}~\bibnamefont {Wu}}, \bibinfo {author} {\bibfnamefont {L.}~\bibnamefont {Li}}, \bibinfo {author} {\bibfnamefont {A.}~\bibnamefont {Basit}}, \bibinfo {author} {\bibfnamefont {Y.}~\bibnamefont {Wei}}, \bibinfo {author} {\bibfnamefont {S.}~\bibnamefont {Chen}}, \bibinfo {author} {\bibfnamefont {P.}~\bibnamefont {Grelu}},\ and\ \bibinfo {author} {\bibfnamefont {Z.}~\bibnamefont {Ni}},\ }\bibfield  {title} {\bibinfo {title} {Non-hermitian skin effects in two- and three-dimensional intertwined tight-binding lattices},\ }\href {https://doi.org/10.1103/PhysRevB.109.205135} {\bibfield  {journal} {\bibinfo  {journal} {Phys. Rev. B}\ }\textbf {\bibinfo {volume} {109}},\ \bibinfo {pages} {205135} (\bibinfo {year} {2024})}\BibitemShut {NoStop}%
\bibitem [{\citenamefont {Gliozzi}\ \emph {et~al.}(2024)\citenamefont {Gliozzi}, \citenamefont {De~Tomasi},\ and\ \citenamefont {Hughes}}]{Gliozzi2024}%
  \BibitemOpen
  \bibfield  {author} {\bibinfo {author} {\bibfnamefont {J.}~\bibnamefont {Gliozzi}}, \bibinfo {author} {\bibfnamefont {G.}~\bibnamefont {De~Tomasi}},\ and\ \bibinfo {author} {\bibfnamefont {T.~L.}\ \bibnamefont {Hughes}},\ }\bibfield  {title} {\bibinfo {title} {Many-body non-hermitian skin effect for multipoles},\ }\href {https://doi.org/10.1103/PhysRevLett.133.136503} {\bibfield  {journal} {\bibinfo  {journal} {Phys. Rev. Lett.}\ }\textbf {\bibinfo {volume} {133}},\ \bibinfo {pages} {136503} (\bibinfo {year} {2024})}\BibitemShut {NoStop}%
\bibitem [{\citenamefont {Okugawa}\ \emph {et~al.}(2020)\citenamefont {Okugawa}, \citenamefont {Takahashi},\ and\ \citenamefont {Yokomizo}}]{Okugawa2020}%
  \BibitemOpen
  \bibfield  {author} {\bibinfo {author} {\bibfnamefont {R.}~\bibnamefont {Okugawa}}, \bibinfo {author} {\bibfnamefont {R.}~\bibnamefont {Takahashi}},\ and\ \bibinfo {author} {\bibfnamefont {K.}~\bibnamefont {Yokomizo}},\ }\bibfield  {title} {\bibinfo {title} {Second-order topological non-hermitian skin effects},\ }\href {https://doi.org/10.1103/PhysRevB.102.241202} {\bibfield  {journal} {\bibinfo  {journal} {Phys. Rev. B}\ }\textbf {\bibinfo {volume} {102}},\ \bibinfo {pages} {241202} (\bibinfo {year} {2020})}\BibitemShut {NoStop}%
\bibitem [{\citenamefont {Yokomizo}\ and\ \citenamefont {Murakami}(2021)}]{Yokomizo2021}%
  \BibitemOpen
  \bibfield  {author} {\bibinfo {author} {\bibfnamefont {K.}~\bibnamefont {Yokomizo}}\ and\ \bibinfo {author} {\bibfnamefont {S.}~\bibnamefont {Murakami}},\ }\bibfield  {title} {\bibinfo {title} {Scaling rule for the critical non-hermitian skin effect},\ }\href {https://doi.org/10.1103/PhysRevB.104.165117} {\bibfield  {journal} {\bibinfo  {journal} {Phys. Rev. B}\ }\textbf {\bibinfo {volume} {104}},\ \bibinfo {pages} {165117} (\bibinfo {year} {2021})}\BibitemShut {NoStop}%
\bibitem [{\citenamefont {Longhi}(2022)}]{Longhi2022}%
  \BibitemOpen
  \bibfield  {author} {\bibinfo {author} {\bibfnamefont {S.}~\bibnamefont {Longhi}},\ }\bibfield  {title} {\bibinfo {title} {Non-hermitian skin effect and self-acceleration},\ }\href {https://doi.org/10.1103/PhysRevB.105.245143} {\bibfield  {journal} {\bibinfo  {journal} {Phys. Rev. B}\ }\textbf {\bibinfo {volume} {105}},\ \bibinfo {pages} {245143} (\bibinfo {year} {2022})}\BibitemShut {NoStop}%
\bibitem [{\citenamefont {Song}\ \emph {et~al.}(2019)\citenamefont {Song}, \citenamefont {Yao},\ and\ \citenamefont {Wang}}]{Song2019}%
  \BibitemOpen
  \bibfield  {author} {\bibinfo {author} {\bibfnamefont {F.}~\bibnamefont {Song}}, \bibinfo {author} {\bibfnamefont {S.}~\bibnamefont {Yao}},\ and\ \bibinfo {author} {\bibfnamefont {Z.}~\bibnamefont {Wang}},\ }\bibfield  {title} {\bibinfo {title} {Non-hermitian skin effect and chiral damping in open quantum systems},\ }\href {https://doi.org/10.1103/PhysRevLett.123.170401} {\bibfield  {journal} {\bibinfo  {journal} {Phys. Rev. Lett.}\ }\textbf {\bibinfo {volume} {123}},\ \bibinfo {pages} {170401} (\bibinfo {year} {2019})}\BibitemShut {NoStop}%
\bibitem [{\citenamefont {Li}\ \emph {et~al.}(2020)\citenamefont {Li}, \citenamefont {Lee},\ and\ \citenamefont {Gong}}]{Li2020}%
  \BibitemOpen
  \bibfield  {author} {\bibinfo {author} {\bibfnamefont {L.}~\bibnamefont {Li}}, \bibinfo {author} {\bibfnamefont {C.~H.}\ \bibnamefont {Lee}},\ and\ \bibinfo {author} {\bibfnamefont {J.}~\bibnamefont {Gong}},\ }\bibfield  {title} {\bibinfo {title} {Topological switch for non-hermitian skin effect in cold-atom systems with loss},\ }\href {https://doi.org/10.1103/PhysRevLett.124.250402} {\bibfield  {journal} {\bibinfo  {journal} {Phys. Rev. Lett.}\ }\textbf {\bibinfo {volume} {124}},\ \bibinfo {pages} {250402} (\bibinfo {year} {2020})}\BibitemShut {NoStop}%
\bibitem [{\citenamefont {Sun}\ \emph {et~al.}(2021)\citenamefont {Sun}, \citenamefont {Zhu},\ and\ \citenamefont {Hughes}}]{Sun2021b}%
  \BibitemOpen
  \bibfield  {author} {\bibinfo {author} {\bibfnamefont {X.-Q.}\ \bibnamefont {Sun}}, \bibinfo {author} {\bibfnamefont {P.}~\bibnamefont {Zhu}},\ and\ \bibinfo {author} {\bibfnamefont {T.~L.}\ \bibnamefont {Hughes}},\ }\bibfield  {title} {\bibinfo {title} {Geometric response and disclination-induced skin effects in non-hermitian systems},\ }\href {https://doi.org/10.1103/PhysRevLett.127.066401} {\bibfield  {journal} {\bibinfo  {journal} {Phys. Rev. Lett.}\ }\textbf {\bibinfo {volume} {127}},\ \bibinfo {pages} {066401} (\bibinfo {year} {2021})}\BibitemShut {NoStop}%
\bibitem [{\citenamefont {Li}\ \emph {et~al.}(2023)\citenamefont {Li}, \citenamefont {Trauzettel}, \citenamefont {Neupert},\ and\ \citenamefont {Zhang}}]{Li2023e}%
  \BibitemOpen
  \bibfield  {author} {\bibinfo {author} {\bibfnamefont {C.-A.}\ \bibnamefont {Li}}, \bibinfo {author} {\bibfnamefont {B.}~\bibnamefont {Trauzettel}}, \bibinfo {author} {\bibfnamefont {T.}~\bibnamefont {Neupert}},\ and\ \bibinfo {author} {\bibfnamefont {S.-B.}\ \bibnamefont {Zhang}},\ }\bibfield  {title} {\bibinfo {title} {Enhancement of second-order non-hermitian skin effect by magnetic fields},\ }\href {https://doi.org/10.1103/PhysRevLett.131.116601} {\bibfield  {journal} {\bibinfo  {journal} {Phys. Rev. Lett.}\ }\textbf {\bibinfo {volume} {131}},\ \bibinfo {pages} {116601} (\bibinfo {year} {2023})}\BibitemShut {NoStop}%
\bibitem [{\citenamefont {Liu}\ \emph {et~al.}(2024)\citenamefont {Liu}, \citenamefont {Mandal}, \citenamefont {Zhou}, \citenamefont {Xi}, \citenamefont {Banerjee}, \citenamefont {Hu}, \citenamefont {Wei}, \citenamefont {Wang}, \citenamefont {Wang}, \citenamefont {Gao}, \citenamefont {Chen}, \citenamefont {Yang}, \citenamefont {Chong},\ and\ \citenamefont {Zhang}}]{Liu2024a}%
  \BibitemOpen
  \bibfield  {author} {\bibinfo {author} {\bibfnamefont {G.-G.}\ \bibnamefont {Liu}}, \bibinfo {author} {\bibfnamefont {S.}~\bibnamefont {Mandal}}, \bibinfo {author} {\bibfnamefont {P.}~\bibnamefont {Zhou}}, \bibinfo {author} {\bibfnamefont {X.}~\bibnamefont {Xi}}, \bibinfo {author} {\bibfnamefont {R.}~\bibnamefont {Banerjee}}, \bibinfo {author} {\bibfnamefont {Y.-H.}\ \bibnamefont {Hu}}, \bibinfo {author} {\bibfnamefont {M.}~\bibnamefont {Wei}}, \bibinfo {author} {\bibfnamefont {M.}~\bibnamefont {Wang}}, \bibinfo {author} {\bibfnamefont {Q.}~\bibnamefont {Wang}}, \bibinfo {author} {\bibfnamefont {Z.}~\bibnamefont {Gao}}, \bibinfo {author} {\bibfnamefont {H.}~\bibnamefont {Chen}}, \bibinfo {author} {\bibfnamefont {Y.}~\bibnamefont {Yang}}, \bibinfo {author} {\bibfnamefont {Y.}~\bibnamefont {Chong}},\ and\ \bibinfo {author} {\bibfnamefont {B.}~\bibnamefont {Zhang}},\ }\bibfield  {title} {\bibinfo {title} {Localization of chiral edge states by the non-hermitian skin effect},\ }\href {https://doi.org/10.1103/PhysRevLett.132.113802} {\bibfield  {journal} {\bibinfo  {journal} {Phys. Rev. Lett.}\ }\textbf {\bibinfo {volume} {132}},\ \bibinfo {pages} {113802} (\bibinfo {year} {2024})}\BibitemShut {NoStop}%
\bibitem [{\citenamefont {Shimomura}\ and\ \citenamefont {Sato}(2024)}]{Shimomura2024}%
  \BibitemOpen
  \bibfield  {author} {\bibinfo {author} {\bibfnamefont {K.}~\bibnamefont {Shimomura}}\ and\ \bibinfo {author} {\bibfnamefont {M.}~\bibnamefont {Sato}},\ }\bibfield  {title} {\bibinfo {title} {General criterion for non-hermitian skin effects and application: Fock space skin effects in many-body systems},\ }\href {https://doi.org/10.1103/PhysRevLett.133.136502} {\bibfield  {journal} {\bibinfo  {journal} {Phys. Rev. Lett.}\ }\textbf {\bibinfo {volume} {133}},\ \bibinfo {pages} {136502} (\bibinfo {year} {2024})}\BibitemShut {NoStop}%
\bibitem [{\citenamefont {Kawabata}\ \emph {et~al.}(2023)\citenamefont {Kawabata}, \citenamefont {Numasawa},\ and\ \citenamefont {Ryu}}]{Kawabata2023}%
  \BibitemOpen
  \bibfield  {author} {\bibinfo {author} {\bibfnamefont {K.}~\bibnamefont {Kawabata}}, \bibinfo {author} {\bibfnamefont {T.}~\bibnamefont {Numasawa}},\ and\ \bibinfo {author} {\bibfnamefont {S.}~\bibnamefont {Ryu}},\ }\bibfield  {title} {\bibinfo {title} {Entanglement phase transition induced by the non-{H}ermitian skin effect},\ }\href {https://doi.org/10.1103/PhysRevX.13.021007} {\bibfield  {journal} {\bibinfo  {journal} {Phys. Rev. X}\ }\textbf {\bibinfo {volume} {13}},\ \bibinfo {pages} {021007} (\bibinfo {year} {2023})}\BibitemShut {NoStop}%
\bibitem [{\citenamefont {Zhou}\ \emph {et~al.}(2025)\citenamefont {Zhou}, \citenamefont {Nie}, \citenamefont {Hu}, \citenamefont {Wang}, \citenamefont {Wu}, \citenamefont {He},\ and\ \citenamefont {Deng}}]{Zhou2025}%
  \BibitemOpen
  \bibfield  {author} {\bibinfo {author} {\bibfnamefont {D.}~\bibnamefont {Zhou}}, \bibinfo {author} {\bibfnamefont {L.}~\bibnamefont {Nie}}, \bibinfo {author} {\bibfnamefont {R.}~\bibnamefont {Hu}}, \bibinfo {author} {\bibfnamefont {X.}~\bibnamefont {Wang}}, \bibinfo {author} {\bibfnamefont {J.}~\bibnamefont {Wu}}, \bibinfo {author} {\bibfnamefont {Z.}~\bibnamefont {He}},\ and\ \bibinfo {author} {\bibfnamefont {K.}~\bibnamefont {Deng}},\ }\bibfield  {title} {\bibinfo {title} {Observation of anomalous non-hermitian skin effect in electric circuits},\ }\href {https://doi.org/10.1103/PhysRevB.111.224104} {\bibfield  {journal} {\bibinfo  {journal} {Phys. Rev. B}\ }\textbf {\bibinfo {volume} {111}},\ \bibinfo {pages} {224104} (\bibinfo {year} {2025})}\BibitemShut {NoStop}%
\bibitem [{\citenamefont {Xu}\ \emph {et~al.}(2025)\citenamefont {Xu}, \citenamefont {Guan},\ and\ \citenamefont {Xu}}]{Xu2025}%
  \BibitemOpen
  \bibfield  {author} {\bibinfo {author} {\bibfnamefont {C.}~\bibnamefont {Xu}}, \bibinfo {author} {\bibfnamefont {Z.}~\bibnamefont {Guan}},\ and\ \bibinfo {author} {\bibfnamefont {H.}~\bibnamefont {Xu}},\ }\bibfield  {title} {\bibinfo {title} {Controllable suppression of non-hermitian skin effects},\ }\href {https://doi.org/10.1103/PhysRevB.111.024201} {\bibfield  {journal} {\bibinfo  {journal} {Phys. Rev. B}\ }\textbf {\bibinfo {volume} {111}},\ \bibinfo {pages} {024201} (\bibinfo {year} {2025})}\BibitemShut {NoStop}%
\bibitem [{\citenamefont {Wang}\ \emph {et~al.}(2023)\citenamefont {Wang}, \citenamefont {Suthar}, \citenamefont {Jen}, \citenamefont {Hsu},\ and\ \citenamefont {You}}]{Wang2023a}%
  \BibitemOpen
  \bibfield  {author} {\bibinfo {author} {\bibfnamefont {Y.-C.}\ \bibnamefont {Wang}}, \bibinfo {author} {\bibfnamefont {K.}~\bibnamefont {Suthar}}, \bibinfo {author} {\bibfnamefont {H.~H.}\ \bibnamefont {Jen}}, \bibinfo {author} {\bibfnamefont {Y.-T.}\ \bibnamefont {Hsu}},\ and\ \bibinfo {author} {\bibfnamefont {J.-S.}\ \bibnamefont {You}},\ }\bibfield  {title} {\bibinfo {title} {Non-hermitian skin effects on thermal and many-body localized phases},\ }\href {https://doi.org/10.1103/PhysRevB.107.L220205} {\bibfield  {journal} {\bibinfo  {journal} {Phys. Rev. B}\ }\textbf {\bibinfo {volume} {107}},\ \bibinfo {pages} {L220205} (\bibinfo {year} {2023})}\BibitemShut {NoStop}%
\bibitem [{\citenamefont {Kunst}\ \emph {et~al.}(2018)\citenamefont {Kunst}, \citenamefont {Edvardsson}, \citenamefont {Budich},\ and\ \citenamefont {Bergholtz}}]{Kunst2018}%
  \BibitemOpen
  \bibfield  {author} {\bibinfo {author} {\bibfnamefont {F.~K.}\ \bibnamefont {Kunst}}, \bibinfo {author} {\bibfnamefont {E.}~\bibnamefont {Edvardsson}}, \bibinfo {author} {\bibfnamefont {J.~C.}\ \bibnamefont {Budich}},\ and\ \bibinfo {author} {\bibfnamefont {E.~J.}\ \bibnamefont {Bergholtz}},\ }\bibfield  {title} {\bibinfo {title} {Biorthogonal bulk-boundary correspondence in non-{H}ermitian systems},\ }\href {https://doi.org/10.1103/PhysRevLett.121.026808} {\bibfield  {journal} {\bibinfo  {journal} {Phys. Rev. Lett.}\ }\textbf {\bibinfo {volume} {121}},\ \bibinfo {pages} {026808} (\bibinfo {year} {2018})}\BibitemShut {NoStop}%
\bibitem [{\citenamefont {Xiao}\ \emph {et~al.}(2020)\citenamefont {Xiao}, \citenamefont {Deng}, \citenamefont {Wang}, \citenamefont {Zhu}, \citenamefont {Wang}, \citenamefont {Yi},\ and\ \citenamefont {Xue}}]{Xiao2020}%
  \BibitemOpen
  \bibfield  {author} {\bibinfo {author} {\bibfnamefont {L.}~\bibnamefont {Xiao}}, \bibinfo {author} {\bibfnamefont {T.}~\bibnamefont {Deng}}, \bibinfo {author} {\bibfnamefont {K.}~\bibnamefont {Wang}}, \bibinfo {author} {\bibfnamefont {G.}~\bibnamefont {Zhu}}, \bibinfo {author} {\bibfnamefont {Z.}~\bibnamefont {Wang}}, \bibinfo {author} {\bibfnamefont {W.}~\bibnamefont {Yi}},\ and\ \bibinfo {author} {\bibfnamefont {P.}~\bibnamefont {Xue}},\ }\bibfield  {title} {\bibinfo {title} {Non-{H}ermitian bulk-boundary correspondence in quantum dynamics},\ }\href {https://doi.org/10.1038/s41567-020-0836-6} {\bibfield  {journal} {\bibinfo  {journal} {Nat. Phys.}\ }\textbf {\bibinfo {volume} {16}},\ \bibinfo {pages} {761} (\bibinfo {year} {2020})}\BibitemShut {NoStop}%
\bibitem [{\citenamefont {Trifunovic}\ and\ \citenamefont {Brouwer}(2019)}]{Trifunovic2019}%
  \BibitemOpen
  \bibfield  {author} {\bibinfo {author} {\bibfnamefont {L.}~\bibnamefont {Trifunovic}}\ and\ \bibinfo {author} {\bibfnamefont {P.~W.}\ \bibnamefont {Brouwer}},\ }\bibfield  {title} {\bibinfo {title} {Higher-order bulk-boundary correspondence for topological crystalline phases},\ }\href {https://doi.org/10.1103/PhysRevX.9.011012} {\bibfield  {journal} {\bibinfo  {journal} {Phys. Rev. X}\ }\textbf {\bibinfo {volume} {9}},\ \bibinfo {pages} {011012} (\bibinfo {year} {2019})}\BibitemShut {NoStop}%
\bibitem [{\citenamefont {Nakamura}\ \emph {et~al.}(2024)\citenamefont {Nakamura}, \citenamefont {Bessho},\ and\ \citenamefont {Sato}}]{Nakamura2024}%
  \BibitemOpen
  \bibfield  {author} {\bibinfo {author} {\bibfnamefont {D.}~\bibnamefont {Nakamura}}, \bibinfo {author} {\bibfnamefont {T.}~\bibnamefont {Bessho}},\ and\ \bibinfo {author} {\bibfnamefont {M.}~\bibnamefont {Sato}},\ }\bibfield  {title} {\bibinfo {title} {Bulk-boundary correspondence in point-gap topological phases},\ }\href {https://doi.org/10.1103/PhysRevLett.132.136401} {\bibfield  {journal} {\bibinfo  {journal} {Phys. Rev. Lett.}\ }\textbf {\bibinfo {volume} {132}},\ \bibinfo {pages} {136401} (\bibinfo {year} {2024})}\BibitemShut {NoStop}%
\bibitem [{\citenamefont {Yang}\ \emph {et~al.}(2020)\citenamefont {Yang}, \citenamefont {Zhang}, \citenamefont {Fang},\ and\ \citenamefont {Hu}}]{Yang2020}%
  \BibitemOpen
  \bibfield  {author} {\bibinfo {author} {\bibfnamefont {Z.}~\bibnamefont {Yang}}, \bibinfo {author} {\bibfnamefont {K.}~\bibnamefont {Zhang}}, \bibinfo {author} {\bibfnamefont {C.}~\bibnamefont {Fang}},\ and\ \bibinfo {author} {\bibfnamefont {J.}~\bibnamefont {Hu}},\ }\bibfield  {title} {\bibinfo {title} {Non-{H}ermitian bulk-boundary correspondence and auxiliary generalized brillouin zone theory},\ }\href {https://doi.org/10.1103/PhysRevLett.125.226402} {\bibfield  {journal} {\bibinfo  {journal} {Phys. Rev. Lett.}\ }\textbf {\bibinfo {volume} {125}},\ \bibinfo {pages} {226402} (\bibinfo {year} {2020})}\BibitemShut {NoStop}%
\bibitem [{\citenamefont {Rhim}\ \emph {et~al.}(2017)\citenamefont {Rhim}, \citenamefont {Behrends},\ and\ \citenamefont {Bardarson}}]{Rhim2017}%
  \BibitemOpen
  \bibfield  {author} {\bibinfo {author} {\bibfnamefont {J.-W.}\ \bibnamefont {Rhim}}, \bibinfo {author} {\bibfnamefont {J.}~\bibnamefont {Behrends}},\ and\ \bibinfo {author} {\bibfnamefont {J.~H.}\ \bibnamefont {Bardarson}},\ }\bibfield  {title} {\bibinfo {title} {Bulk-boundary correspondence from the intercellular zak phase},\ }\href {https://doi.org/10.1103/PhysRevB.95.035421} {\bibfield  {journal} {\bibinfo  {journal} {Phys. Rev. B}\ }\textbf {\bibinfo {volume} {95}},\ \bibinfo {pages} {035421} (\bibinfo {year} {2017})}\BibitemShut {NoStop}%
\bibitem [{\citenamefont {Tamura}\ \emph {et~al.}(2021)\citenamefont {Tamura}, \citenamefont {Hoshino},\ and\ \citenamefont {Tanaka}}]{Tamura2021}%
  \BibitemOpen
  \bibfield  {author} {\bibinfo {author} {\bibfnamefont {S.}~\bibnamefont {Tamura}}, \bibinfo {author} {\bibfnamefont {S.}~\bibnamefont {Hoshino}},\ and\ \bibinfo {author} {\bibfnamefont {Y.}~\bibnamefont {Tanaka}},\ }\bibfield  {title} {\bibinfo {title} {Generalization of spectral bulk-boundary correspondence},\ }\href {https://doi.org/10.1103/PhysRevB.104.165125} {\bibfield  {journal} {\bibinfo  {journal} {Phys. Rev. B}\ }\textbf {\bibinfo {volume} {104}},\ \bibinfo {pages} {165125} (\bibinfo {year} {2021})}\BibitemShut {NoStop}%
\bibitem [{\citenamefont {Zhang}\ \emph {et~al.}(2020)\citenamefont {Zhang}, \citenamefont {Yang},\ and\ \citenamefont {Fang}}]{Zhang2020a}%
  \BibitemOpen
  \bibfield  {author} {\bibinfo {author} {\bibfnamefont {K.}~\bibnamefont {Zhang}}, \bibinfo {author} {\bibfnamefont {Z.}~\bibnamefont {Yang}},\ and\ \bibinfo {author} {\bibfnamefont {C.}~\bibnamefont {Fang}},\ }\bibfield  {title} {\bibinfo {title} {Correspondence between winding numbers and skin modes in non-{H}ermitian systems},\ }\href {https://doi.org/10.1103/PhysRevLett.125.126402} {\bibfield  {journal} {\bibinfo  {journal} {Phys. Rev. Lett.}\ }\textbf {\bibinfo {volume} {125}},\ \bibinfo {pages} {126402} (\bibinfo {year} {2020})}\BibitemShut {NoStop}%
\bibitem [{\citenamefont {Yokomizo}\ and\ \citenamefont {Murakami}(2019)}]{Yokomizo2019}%
  \BibitemOpen
  \bibfield  {author} {\bibinfo {author} {\bibfnamefont {K.}~\bibnamefont {Yokomizo}}\ and\ \bibinfo {author} {\bibfnamefont {S.}~\bibnamefont {Murakami}},\ }\bibfield  {title} {\bibinfo {title} {Non-{B}loch band theory of non-{H}ermitian systems},\ }\href {https://doi.org/10.1103/PhysRevLett.123.066404} {\bibfield  {journal} {\bibinfo  {journal} {Phys. Rev. Lett.}\ }\textbf {\bibinfo {volume} {123}},\ \bibinfo {pages} {066404} (\bibinfo {year} {2019})}\BibitemShut {NoStop}%
\bibitem [{\citenamefont {Kawabata}\ \emph {et~al.}(2019)\citenamefont {Kawabata}, \citenamefont {Shiozaki}, \citenamefont {Ueda},\ and\ \citenamefont {Sato}}]{Kawabata2019}%
  \BibitemOpen
  \bibfield  {author} {\bibinfo {author} {\bibfnamefont {K.}~\bibnamefont {Kawabata}}, \bibinfo {author} {\bibfnamefont {K.}~\bibnamefont {Shiozaki}}, \bibinfo {author} {\bibfnamefont {M.}~\bibnamefont {Ueda}},\ and\ \bibinfo {author} {\bibfnamefont {M.}~\bibnamefont {Sato}},\ }\bibfield  {title} {\bibinfo {title} {Symmetry and topology in non-{H}ermitian physics},\ }\href {https://doi.org/10.1103/PhysRevX.9.041015} {\bibfield  {journal} {\bibinfo  {journal} {Phys. Rev. X}\ }\textbf {\bibinfo {volume} {9}},\ \bibinfo {pages} {041015} (\bibinfo {year} {2019})}\BibitemShut {NoStop}%
\bibitem [{\citenamefont {Borgnia}\ \emph {et~al.}(2020)\citenamefont {Borgnia}, \citenamefont {Kruchkov},\ and\ \citenamefont {Slager}}]{Borgnia2020}%
  \BibitemOpen
  \bibfield  {author} {\bibinfo {author} {\bibfnamefont {D.~S.}\ \bibnamefont {Borgnia}}, \bibinfo {author} {\bibfnamefont {A.~J.}\ \bibnamefont {Kruchkov}},\ and\ \bibinfo {author} {\bibfnamefont {R.-J.}\ \bibnamefont {Slager}},\ }\bibfield  {title} {\bibinfo {title} {Non-{H}ermitian boundary modes and topology},\ }\href {https://doi.org/10.1103/PhysRevLett.124.056802} {\bibfield  {journal} {\bibinfo  {journal} {Phys. Rev. Lett.}\ }\textbf {\bibinfo {volume} {124}},\ \bibinfo {pages} {056802} (\bibinfo {year} {2020})}\BibitemShut {NoStop}%
\bibitem [{\citenamefont {Kaneshiro}\ and\ \citenamefont {Peters}(2025)}]{Kaneshiro2025}%
  \BibitemOpen
  \bibfield  {author} {\bibinfo {author} {\bibfnamefont {S.}~\bibnamefont {Kaneshiro}}\ and\ \bibinfo {author} {\bibfnamefont {R.}~\bibnamefont {Peters}},\ }\bibfield  {title} {\bibinfo {title} {Symplectic-amoeba formulation of the non-bloch band theory for one-dimensional two-band systems},\ }\href {https://doi.org/10.1103/5s1z-5t9r} {\bibfield  {journal} {\bibinfo  {journal} {Phys. Rev. B}\ }\textbf {\bibinfo {volume} {112}},\ \bibinfo {pages} {075408} (\bibinfo {year} {2025})}\BibitemShut {NoStop}%
\bibitem [{\citenamefont {Yokomizo}\ \emph {et~al.}(2024)\citenamefont {Yokomizo}, \citenamefont {Yoda},\ and\ \citenamefont {Ashida}}]{Yokomizo2024}%
  \BibitemOpen
  \bibfield  {author} {\bibinfo {author} {\bibfnamefont {K.}~\bibnamefont {Yokomizo}}, \bibinfo {author} {\bibfnamefont {T.}~\bibnamefont {Yoda}},\ and\ \bibinfo {author} {\bibfnamefont {Y.}~\bibnamefont {Ashida}},\ }\bibfield  {title} {\bibinfo {title} {Non-bloch band theory of generalized eigenvalue problems},\ }\href {https://doi.org/10.1103/PhysRevB.109.115115} {\bibfield  {journal} {\bibinfo  {journal} {Phys. Rev. B}\ }\textbf {\bibinfo {volume} {109}},\ \bibinfo {pages} {115115} (\bibinfo {year} {2024})}\BibitemShut {NoStop}%
\bibitem [{\citenamefont {Yang}\ \emph {et~al.}(2024)\citenamefont {Yang}, \citenamefont {Lu},\ and\ \citenamefont {Lu}}]{Yang2024}%
  \BibitemOpen
  \bibfield  {author} {\bibinfo {author} {\bibfnamefont {Z.}~\bibnamefont {Yang}}, \bibinfo {author} {\bibfnamefont {C.}~\bibnamefont {Lu}},\ and\ \bibinfo {author} {\bibfnamefont {X.}~\bibnamefont {Lu}},\ }\bibfield  {title} {\bibinfo {title} {Entanglement entropy on generalized {B}rillouin zone},\ }\href {https://doi.org/10.1103/PhysRevB.110.235127} {\bibfield  {journal} {\bibinfo  {journal} {Phys. Rev. B}\ }\textbf {\bibinfo {volume} {110}},\ \bibinfo {pages} {235127} (\bibinfo {year} {2024})}\BibitemShut {NoStop}%
\bibitem [{\citenamefont {Wang}\ \emph {et~al.}(2024)\citenamefont {Wang}, \citenamefont {Song},\ and\ \citenamefont {Wang}}]{Wang2024b}%
  \BibitemOpen
  \bibfield  {author} {\bibinfo {author} {\bibfnamefont {H.-Y.}\ \bibnamefont {Wang}}, \bibinfo {author} {\bibfnamefont {F.}~\bibnamefont {Song}},\ and\ \bibinfo {author} {\bibfnamefont {Z.}~\bibnamefont {Wang}},\ }\bibfield  {title} {\bibinfo {title} {Amoeba formulation of non-{B}loch band theory in arbitrary dimensions},\ }\href {https://doi.org/10.1103/PhysRevX.14.021011} {\bibfield  {journal} {\bibinfo  {journal} {Phys. Rev. X}\ }\textbf {\bibinfo {volume} {14}},\ \bibinfo {pages} {021011} (\bibinfo {year} {2024})}\BibitemShut {NoStop}%
\bibitem [{\citenamefont {Longhi}(2020{\natexlab{b}})}]{Longhi2020a}%
  \BibitemOpen
  \bibfield  {author} {\bibinfo {author} {\bibfnamefont {S.}~\bibnamefont {Longhi}},\ }\bibfield  {title} {\bibinfo {title} {Non-bloch-band collapse and chiral zener tunneling},\ }\href {https://doi.org/10.1103/PhysRevLett.124.066602} {\bibfield  {journal} {\bibinfo  {journal} {Phys. Rev. Lett.}\ }\textbf {\bibinfo {volume} {124}},\ \bibinfo {pages} {066602} (\bibinfo {year} {2020}{\natexlab{b}})}\BibitemShut {NoStop}%
\bibitem [{\citenamefont {Verma}\ and\ \citenamefont {Park}(2024)}]{Verma2024}%
  \BibitemOpen
  \bibfield  {author} {\bibinfo {author} {\bibfnamefont {S.}~\bibnamefont {Verma}}\ and\ \bibinfo {author} {\bibfnamefont {M.~J.}\ \bibnamefont {Park}},\ }\bibfield  {title} {\bibinfo {title} {Non-bloch band theory of subsymmetry-protected topological phases},\ }\href {https://doi.org/10.1103/PhysRevB.110.035424} {\bibfield  {journal} {\bibinfo  {journal} {Phys. Rev. B}\ }\textbf {\bibinfo {volume} {110}},\ \bibinfo {pages} {035424} (\bibinfo {year} {2024})}\BibitemShut {NoStop}%
\bibitem [{\citenamefont {Xue}\ \emph {et~al.}(2021)\citenamefont {Xue}, \citenamefont {Li}, \citenamefont {Hu}, \citenamefont {Song},\ and\ \citenamefont {Wang}}]{Xue2021}%
  \BibitemOpen
  \bibfield  {author} {\bibinfo {author} {\bibfnamefont {W.-T.}\ \bibnamefont {Xue}}, \bibinfo {author} {\bibfnamefont {M.-R.}\ \bibnamefont {Li}}, \bibinfo {author} {\bibfnamefont {Y.-M.}\ \bibnamefont {Hu}}, \bibinfo {author} {\bibfnamefont {F.}~\bibnamefont {Song}},\ and\ \bibinfo {author} {\bibfnamefont {Z.}~\bibnamefont {Wang}},\ }\bibfield  {title} {\bibinfo {title} {Simple formulas of directional amplification from non-{B}loch band theory},\ }\href {https://doi.org/10.1103/PhysRevB.103.L241408} {\bibfield  {journal} {\bibinfo  {journal} {Phys. Rev. B}\ }\textbf {\bibinfo {volume} {103}},\ \bibinfo {pages} {L241408} (\bibinfo {year} {2021})}\BibitemShut {NoStop}%
\bibitem [{\citenamefont {Kawabata}\ \emph {et~al.}(2020)\citenamefont {Kawabata}, \citenamefont {Okuma},\ and\ \citenamefont {Sato}}]{Kawabata2020}%
  \BibitemOpen
  \bibfield  {author} {\bibinfo {author} {\bibfnamefont {K.}~\bibnamefont {Kawabata}}, \bibinfo {author} {\bibfnamefont {N.}~\bibnamefont {Okuma}},\ and\ \bibinfo {author} {\bibfnamefont {M.}~\bibnamefont {Sato}},\ }\bibfield  {title} {\bibinfo {title} {Non-bloch band theory of non-{H}ermitian {H}amiltonians in the symplectic class},\ }\href {https://doi.org/10.1103/PhysRevB.101.195147} {\bibfield  {journal} {\bibinfo  {journal} {Phys. Rev. B}\ }\textbf {\bibinfo {volume} {101}},\ \bibinfo {pages} {195147} (\bibinfo {year} {2020})}\BibitemShut {NoStop}%
\bibitem [{\citenamefont {Lee}\ \emph {et~al.}(2020)\citenamefont {Lee}, \citenamefont {Li}, \citenamefont {Thomale},\ and\ \citenamefont {Gong}}]{Lee2020}%
  \BibitemOpen
  \bibfield  {author} {\bibinfo {author} {\bibfnamefont {C.~H.}\ \bibnamefont {Lee}}, \bibinfo {author} {\bibfnamefont {L.}~\bibnamefont {Li}}, \bibinfo {author} {\bibfnamefont {R.}~\bibnamefont {Thomale}},\ and\ \bibinfo {author} {\bibfnamefont {J.}~\bibnamefont {Gong}},\ }\bibfield  {title} {\bibinfo {title} {Unraveling non-{H}ermitian pumping: Emergent spectral singularities and anomalous responses},\ }\href {https://doi.org/10.1103/PhysRevB.102.085151} {\bibfield  {journal} {\bibinfo  {journal} {Phys. Rev. B}\ }\textbf {\bibinfo {volume} {102}},\ \bibinfo {pages} {085151} (\bibinfo {year} {2020})}\BibitemShut {NoStop}%
\bibitem [{\citenamefont {Masuda}\ and\ \citenamefont {Nakamura}(2022)}]{Masuda2022}%
  \BibitemOpen
  \bibfield  {author} {\bibinfo {author} {\bibfnamefont {S.}~\bibnamefont {Masuda}}\ and\ \bibinfo {author} {\bibfnamefont {M.}~\bibnamefont {Nakamura}},\ }\bibfield  {title} {\bibinfo {title} {Electronic polarization in non-{B}loch band theory},\ }\href {https://doi.org/10.7566/JPSJ.91.114705} {\bibfield  {journal} {\bibinfo  {journal} {J. Phys. Soc. Jpn.}\ }\textbf {\bibinfo {volume} {91}},\ \bibinfo {pages} {114705} (\bibinfo {year} {2022})}\BibitemShut {NoStop}%
\bibitem [{\citenamefont {King-Smith}\ and\ \citenamefont {Vanderbilt}(1993)}]{KingSmith1993}%
  \BibitemOpen
  \bibfield  {author} {\bibinfo {author} {\bibfnamefont {R.~D.}\ \bibnamefont {King-Smith}}\ and\ \bibinfo {author} {\bibfnamefont {D.}~\bibnamefont {Vanderbilt}},\ }\bibfield  {title} {\bibinfo {title} {Theory of polarization of crystalline solids},\ }\href {https://doi.org/10.1103/PhysRevB.47.1651} {\bibfield  {journal} {\bibinfo  {journal} {Phys. Rev. B}\ }\textbf {\bibinfo {volume} {47}},\ \bibinfo {pages} {1651} (\bibinfo {year} {1993})}\BibitemShut {NoStop}%
\bibitem [{\citenamefont {Resta}(1994)}]{Resta1994}%
  \BibitemOpen
  \bibfield  {author} {\bibinfo {author} {\bibfnamefont {R.}~\bibnamefont {Resta}},\ }\bibfield  {title} {\bibinfo {title} {Macroscopic polarization in crystalline dielectrics: the geometric phase approach},\ }\href {https://doi.org/10.1103/RevModPhys.66.899} {\bibfield  {journal} {\bibinfo  {journal} {Rev. Mod. Phys.}\ }\textbf {\bibinfo {volume} {66}},\ \bibinfo {pages} {899} (\bibinfo {year} {1994})}\BibitemShut {NoStop}%
\bibitem [{\citenamefont {Watanabe}\ and\ \citenamefont {Oshikawa}(2018)}]{Watanabe2018}%
  \BibitemOpen
  \bibfield  {author} {\bibinfo {author} {\bibfnamefont {H.}~\bibnamefont {Watanabe}}\ and\ \bibinfo {author} {\bibfnamefont {M.}~\bibnamefont {Oshikawa}},\ }\bibfield  {title} {\bibinfo {title} {Inequivalent berry phases for the bulk polarization},\ }\href {https://doi.org/10.1103/PhysRevX.8.021065} {\bibfield  {journal} {\bibinfo  {journal} {Phys. Rev. X}\ }\textbf {\bibinfo {volume} {8}},\ \bibinfo {pages} {021065} (\bibinfo {year} {2018})}\BibitemShut {NoStop}%
\bibitem [{\citenamefont {Zak}(1989)}]{Zak1989}%
  \BibitemOpen
  \bibfield  {author} {\bibinfo {author} {\bibfnamefont {J.}~\bibnamefont {Zak}},\ }\bibfield  {title} {\bibinfo {title} {Berry's phase for energy bands in solids},\ }\href {https://doi.org/10.1103/PhysRevLett.62.2747} {\bibfield  {journal} {\bibinfo  {journal} {Phys. Rev. Lett.}\ }\textbf {\bibinfo {volume} {62}},\ \bibinfo {pages} {2747} (\bibinfo {year} {1989})}\BibitemShut {NoStop}%
\bibitem [{\citenamefont {Xiao}\ \emph {et~al.}(2010)\citenamefont {Xiao}, \citenamefont {Chang},\ and\ \citenamefont {Niu}}]{Xiao2010}%
  \BibitemOpen
  \bibfield  {author} {\bibinfo {author} {\bibfnamefont {D.}~\bibnamefont {Xiao}}, \bibinfo {author} {\bibfnamefont {M.-C.}\ \bibnamefont {Chang}},\ and\ \bibinfo {author} {\bibfnamefont {Q.}~\bibnamefont {Niu}},\ }\bibfield  {title} {\bibinfo {title} {Berry phase effects on electronic properties},\ }\href {https://doi.org/10.1103/RevModPhys.82.1959} {\bibfield  {journal} {\bibinfo  {journal} {Rev. Mod. Phys.}\ }\textbf {\bibinfo {volume} {82}},\ \bibinfo {pages} {1959} (\bibinfo {year} {2010})}\BibitemShut {NoStop}%
\bibitem [{\citenamefont {Li}\ and\ \citenamefont {Haldane}(2008)}]{Li2008}%
  \BibitemOpen
  \bibfield  {author} {\bibinfo {author} {\bibfnamefont {H.}~\bibnamefont {Li}}\ and\ \bibinfo {author} {\bibfnamefont {F.~D.~M.}\ \bibnamefont {Haldane}},\ }\bibfield  {title} {\bibinfo {title} {Entanglement spectrum as a generalization of entanglement entropy: Identification of topological order in non-abelian fractional quantum hall effect states},\ }\href {https://doi.org/10.1103/PhysRevLett.101.010504} {\bibfield  {journal} {\bibinfo  {journal} {Phys. Rev. Lett.}\ }\textbf {\bibinfo {volume} {101}},\ \bibinfo {pages} {010504} (\bibinfo {year} {2008})}\BibitemShut {NoStop}%
\bibitem [{\citenamefont {Alexandradinata}\ \emph {et~al.}(2011)\citenamefont {Alexandradinata}, \citenamefont {Hughes},\ and\ \citenamefont {Bernevig}}]{Alexandradinata2011}%
  \BibitemOpen
  \bibfield  {author} {\bibinfo {author} {\bibfnamefont {A.}~\bibnamefont {Alexandradinata}}, \bibinfo {author} {\bibfnamefont {T.~L.}\ \bibnamefont {Hughes}},\ and\ \bibinfo {author} {\bibfnamefont {B.~A.}\ \bibnamefont {Bernevig}},\ }\bibfield  {title} {\bibinfo {title} {Trace index and spectral flow in the entanglement spectrum of topological insulators},\ }\href {https://doi.org/10.1103/PhysRevB.84.195103} {\bibfield  {journal} {\bibinfo  {journal} {Phys. Rev. B}\ }\textbf {\bibinfo {volume} {84}},\ \bibinfo {pages} {195103} (\bibinfo {year} {2011})}\BibitemShut {NoStop}%
\bibitem [{\citenamefont {Herviou}\ \emph {et~al.}(2019)\citenamefont {Herviou}, \citenamefont {Regnault},\ and\ \citenamefont {Bardarson}}]{Loic2019}%
  \BibitemOpen
  \bibfield  {author} {\bibinfo {author} {\bibfnamefont {L.}~\bibnamefont {Herviou}}, \bibinfo {author} {\bibfnamefont {N.}~\bibnamefont {Regnault}},\ and\ \bibinfo {author} {\bibfnamefont {J.~H.}\ \bibnamefont {Bardarson}},\ }\bibfield  {title} {\bibinfo {title} {{Entanglement spectrum and symmetries in non-{H}ermitian fermionic non-interacting models}},\ }\href {https://doi.org/10.21468/SciPostPhys.7.5.069} {\bibfield  {journal} {\bibinfo  {journal} {SciPost Phys.}\ }\textbf {\bibinfo {volume} {7}},\ \bibinfo {pages} {069} (\bibinfo {year} {2019})}\BibitemShut {NoStop}%
\bibitem [{\citenamefont {Ortega-Taberner}\ \emph {et~al.}(2022)\citenamefont {Ortega-Taberner}, \citenamefont {R\o{}dland},\ and\ \citenamefont {Hermanns}}]{OrtegaTaberner2022}%
  \BibitemOpen
  \bibfield  {author} {\bibinfo {author} {\bibfnamefont {C.}~\bibnamefont {Ortega-Taberner}}, \bibinfo {author} {\bibfnamefont {L.}~\bibnamefont {R\o{}dland}},\ and\ \bibinfo {author} {\bibfnamefont {M.}~\bibnamefont {Hermanns}},\ }\bibfield  {title} {\bibinfo {title} {Polarization and entanglement spectrum in non-{H}ermitian systems},\ }\href {https://doi.org/10.1103/PhysRevB.105.075103} {\bibfield  {journal} {\bibinfo  {journal} {Phys. Rev. B}\ }\textbf {\bibinfo {volume} {105}},\ \bibinfo {pages} {075103} (\bibinfo {year} {2022})}\BibitemShut {NoStop}%
\bibitem [{\citenamefont {A}\ and\ \citenamefont {B}(2006)}]{Bottcher2006}%
  \BibitemOpen
  \bibfield  {author} {\bibinfo {author} {\bibfnamefont {B.}~\bibnamefont {A}}\ and\ \bibinfo {author} {\bibfnamefont {S.}~\bibnamefont {B}},\ }\href@noop {} {\emph {\bibinfo {title} {Analysis of Toeplitz operators}}}\ (\bibinfo  {publisher} {Berlin, Heidelberg: Springer Berlin Heidelberg},\ \bibinfo {year} {2006})\BibitemShut {NoStop}%
\bibitem [{\citenamefont {Resta}(1998)}]{Resta1998}%
  \BibitemOpen
  \bibfield  {author} {\bibinfo {author} {\bibfnamefont {R.}~\bibnamefont {Resta}},\ }\bibfield  {title} {\bibinfo {title} {Quantum-mechanical position operator in extended systems},\ }\href {https://doi.org/10.1103/PhysRevLett.80.1800} {\bibfield  {journal} {\bibinfo  {journal} {Phys. Rev. Lett.}\ }\textbf {\bibinfo {volume} {80}},\ \bibinfo {pages} {1800} (\bibinfo {year} {1998})}\BibitemShut {NoStop}%
\bibitem [{\citenamefont {Ortega-Taberner}\ and\ \citenamefont {Hermanns}(2021)}]{OrtegaTaberner2021}%
  \BibitemOpen
  \bibfield  {author} {\bibinfo {author} {\bibfnamefont {C.}~\bibnamefont {Ortega-Taberner}}\ and\ \bibinfo {author} {\bibfnamefont {M.}~\bibnamefont {Hermanns}},\ }\bibfield  {title} {\bibinfo {title} {Relation of the entanglement spectrum to the bulk polarization},\ }\href {https://doi.org/10.1103/PhysRevB.103.195132} {\bibfield  {journal} {\bibinfo  {journal} {Phys. Rev. B}\ }\textbf {\bibinfo {volume} {103}},\ \bibinfo {pages} {195132} (\bibinfo {year} {2021})}\BibitemShut {NoStop}%
\bibitem [{\citenamefont {Kohn}(1996)}]{Kohn1996}%
  \BibitemOpen
  \bibfield  {author} {\bibinfo {author} {\bibfnamefont {W.}~\bibnamefont {Kohn}},\ }\bibfield  {title} {\bibinfo {title} {Density functional and density matrix method scaling linearly with the number of atoms},\ }\href {https://doi.org/10.1103/PhysRevLett.76.3168} {\bibfield  {journal} {\bibinfo  {journal} {Phys. Rev. Lett.}\ }\textbf {\bibinfo {volume} {76}},\ \bibinfo {pages} {3168} (\bibinfo {year} {1996})}\BibitemShut {NoStop}%
\bibitem [{\citenamefont {Hatano}\ and\ \citenamefont {Nelson}(1996)}]{Hatano1996}%
  \BibitemOpen
  \bibfield  {author} {\bibinfo {author} {\bibfnamefont {N.}~\bibnamefont {Hatano}}\ and\ \bibinfo {author} {\bibfnamefont {D.~R.}\ \bibnamefont {Nelson}},\ }\bibfield  {title} {\bibinfo {title} {Localization transitions in non-{H}ermitian quantum mechanics},\ }\href {https://doi.org/10.1103/PhysRevLett.77.570} {\bibfield  {journal} {\bibinfo  {journal} {Phys. Rev. Lett.}\ }\textbf {\bibinfo {volume} {77}},\ \bibinfo {pages} {570} (\bibinfo {year} {1996})}\BibitemShut {NoStop}%
\bibitem [{\citenamefont {Ezawa}(2019)}]{Ezawa2019}%
  \BibitemOpen
  \bibfield  {author} {\bibinfo {author} {\bibfnamefont {M.}~\bibnamefont {Ezawa}},\ }\bibfield  {title} {\bibinfo {title} {Non-{H}ermitian boundary and interface states in nonreciprocal higher-order topological metals and electrical circuits},\ }\href {https://doi.org/10.1103/PhysRevB.99.121411} {\bibfield  {journal} {\bibinfo  {journal} {Phys. Rev. B}\ }\textbf {\bibinfo {volume} {99}},\ \bibinfo {pages} {121411} (\bibinfo {year} {2019})}\BibitemShut {NoStop}%
\bibitem [{\citenamefont {Chen}\ \emph {et~al.}(2024)\citenamefont {Chen}, \citenamefont {Sun}, \citenamefont {Wang}, \citenamefont {Jiang}, \citenamefont {Zhu},\ and\ \citenamefont {Zhou}}]{Chen2024a}%
  \BibitemOpen
  \bibfield  {author} {\bibinfo {author} {\bibfnamefont {X.}~\bibnamefont {Chen}}, \bibinfo {author} {\bibfnamefont {J.}~\bibnamefont {Sun}}, \bibinfo {author} {\bibfnamefont {X.}~\bibnamefont {Wang}}, \bibinfo {author} {\bibfnamefont {H.}~\bibnamefont {Jiang}}, \bibinfo {author} {\bibfnamefont {D.}~\bibnamefont {Zhu}},\ and\ \bibinfo {author} {\bibfnamefont {X.}~\bibnamefont {Zhou}},\ }\bibfield  {title} {\bibinfo {title} {Deep learning for the design of non-{H}ermitian topolectrical circuits},\ }\href {https://doi.org/10.1103/PhysRevB.109.094103} {\bibfield  {journal} {\bibinfo  {journal} {Phys. Rev. B}\ }\textbf {\bibinfo {volume} {109}},\ \bibinfo {pages} {094103} (\bibinfo {year} {2024})}\BibitemShut {NoStop}%
\bibitem [{\citenamefont {Sahin}\ \emph {et~al.}(2025)\citenamefont {Sahin}, \citenamefont {Jalil},\ and\ \citenamefont {Lee}}]{Sahin2025}%
  \BibitemOpen
  \bibfield  {author} {\bibinfo {author} {\bibfnamefont {H.}~\bibnamefont {Sahin}}, \bibinfo {author} {\bibfnamefont {M.~B.~A.}\ \bibnamefont {Jalil}},\ and\ \bibinfo {author} {\bibfnamefont {C.~H.}\ \bibnamefont {Lee}},\ }\href@noop {} {\bibinfo {title} {Topolectrical circuits $-$ recent experimental advances and developments}} (\bibinfo {year} {2025}),\ \Eprint {https://arxiv.org/abs/2502.18563} {arXiv:2502.18563 [cond-mat]} \BibitemShut {NoStop}%
\bibitem [{\citenamefont {Yuan}\ \emph {et~al.}(2023)\citenamefont {Yuan}, \citenamefont {Zhang}, \citenamefont {Zhou}, \citenamefont {Wang}, \citenamefont {Pan}, \citenamefont {Feng}, \citenamefont {Sun},\ and\ \citenamefont {Zhang}}]{Yuan2023}%
  \BibitemOpen
  \bibfield  {author} {\bibinfo {author} {\bibfnamefont {H.}~\bibnamefont {Yuan}}, \bibinfo {author} {\bibfnamefont {W.}~\bibnamefont {Zhang}}, \bibinfo {author} {\bibfnamefont {Z.}~\bibnamefont {Zhou}}, \bibinfo {author} {\bibfnamefont {W.}~\bibnamefont {Wang}}, \bibinfo {author} {\bibfnamefont {N.}~\bibnamefont {Pan}}, \bibinfo {author} {\bibfnamefont {Y.}~\bibnamefont {Feng}}, \bibinfo {author} {\bibfnamefont {H.}~\bibnamefont {Sun}},\ and\ \bibinfo {author} {\bibfnamefont {X.}~\bibnamefont {Zhang}},\ }\bibfield  {title} {\bibinfo {title} {Non-{H}ermitian topolectrical circuit sensor with high sensitivity},\ }\href {https://doi.org/10.1002/advs.202301128} {\bibfield  {journal} {\bibinfo  {journal} {Adv. Sci.}\ }\textbf {\bibinfo {volume} {10}},\ \bibinfo {pages} {2301128} (\bibinfo {year} {2023})}\BibitemShut {NoStop}%
\bibitem [{\citenamefont {Ren}\ \emph {et~al.}(2025)\citenamefont {Ren}, \citenamefont {Pan}, \citenamefont {Yao}, \citenamefont {Huo}, \citenamefont {Zheng}, \citenamefont {Hei}, \citenamefont {Qiao},\ and\ \citenamefont {Li}}]{Ren2025}%
  \BibitemOpen
  \bibfield  {author} {\bibinfo {author} {\bibfnamefont {Y.-M.}\ \bibnamefont {Ren}}, \bibinfo {author} {\bibfnamefont {X.-F.}\ \bibnamefont {Pan}}, \bibinfo {author} {\bibfnamefont {X.-Y.}\ \bibnamefont {Yao}}, \bibinfo {author} {\bibfnamefont {X.-W.}\ \bibnamefont {Huo}}, \bibinfo {author} {\bibfnamefont {J.-C.}\ \bibnamefont {Zheng}}, \bibinfo {author} {\bibfnamefont {X.-L.}\ \bibnamefont {Hei}}, \bibinfo {author} {\bibfnamefont {Y.-F.}\ \bibnamefont {Qiao}},\ and\ \bibinfo {author} {\bibfnamefont {P.-B.}\ \bibnamefont {Li}},\ }\bibfield  {title} {\bibinfo {title} {Nonreciprocal interaction and entanglement between two superconducting qubits},\ }\href {https://doi.org/10.1103/213q-sqxf} {\bibfield  {journal} {\bibinfo  {journal} {Phys. Rev. Res.}\ }\textbf {\bibinfo {volume} {7}},\ \bibinfo {pages} {023287} (\bibinfo {year} {2025})}\BibitemShut {NoStop}%
\bibitem [{\citenamefont {Parto}\ \emph {et~al.}(2023)\citenamefont {Parto}, \citenamefont {Leefmans}, \citenamefont {Williams}, \citenamefont {Nori},\ and\ \citenamefont {Marandi}}]{Parto2023}%
  \BibitemOpen
  \bibfield  {author} {\bibinfo {author} {\bibfnamefont {M.}~\bibnamefont {Parto}}, \bibinfo {author} {\bibfnamefont {C.}~\bibnamefont {Leefmans}}, \bibinfo {author} {\bibfnamefont {J.}~\bibnamefont {Williams}}, \bibinfo {author} {\bibfnamefont {F.}~\bibnamefont {Nori}},\ and\ \bibinfo {author} {\bibfnamefont {A.}~\bibnamefont {Marandi}},\ }\bibfield  {title} {\bibinfo {title} {Non-abelian effects in dissipative photonic topological lattices},\ }\href {https://doi.org/10.1038/s41467-023-37065-z} {\bibfield  {journal} {\bibinfo  {journal} {Nat. Commun.}\ }\textbf {\bibinfo {volume} {14}},\ \bibinfo {pages} {1440} (\bibinfo {year} {2023})}\BibitemShut {NoStop}%
\bibitem [{\citenamefont {Gao}\ \emph {et~al.}(2024)\citenamefont {Gao}, \citenamefont {Xu}, \citenamefont {Yang}, \citenamefont {Zwiller},\ and\ \citenamefont {Elshaari}}]{Gao2024}%
  \BibitemOpen
  \bibfield  {author} {\bibinfo {author} {\bibfnamefont {J.}~\bibnamefont {Gao}}, \bibinfo {author} {\bibfnamefont {Z.-S.}\ \bibnamefont {Xu}}, \bibinfo {author} {\bibfnamefont {Z.}~\bibnamefont {Yang}}, \bibinfo {author} {\bibfnamefont {V.}~\bibnamefont {Zwiller}},\ and\ \bibinfo {author} {\bibfnamefont {A.~W.}\ \bibnamefont {Elshaari}},\ }\bibfield  {title} {\bibinfo {title} {Quantum topological photonics with special focus on waveguide systems},\ }\href {https://doi.org/10.1038/s44310-024-00034-5} {\bibfield  {journal} {\bibinfo  {journal} {npj Nanophoton.}\ }\textbf {\bibinfo {volume} {1}},\ \bibinfo {pages} {34} (\bibinfo {year} {2024})}\BibitemShut {NoStop}%
\bibitem [{\citenamefont {Liang}\ \emph {et~al.}(2022)\citenamefont {Liang}, \citenamefont {Xie}, \citenamefont {Dong}, \citenamefont {Li}, \citenamefont {Li}, \citenamefont {Gadway}, \citenamefont {Yi},\ and\ \citenamefont {Yan}}]{Liang2022}%
  \BibitemOpen
  \bibfield  {author} {\bibinfo {author} {\bibfnamefont {Q.}~\bibnamefont {Liang}}, \bibinfo {author} {\bibfnamefont {D.}~\bibnamefont {Xie}}, \bibinfo {author} {\bibfnamefont {Z.}~\bibnamefont {Dong}}, \bibinfo {author} {\bibfnamefont {H.}~\bibnamefont {Li}}, \bibinfo {author} {\bibfnamefont {H.}~\bibnamefont {Li}}, \bibinfo {author} {\bibfnamefont {B.}~\bibnamefont {Gadway}}, \bibinfo {author} {\bibfnamefont {W.}~\bibnamefont {Yi}},\ and\ \bibinfo {author} {\bibfnamefont {B.}~\bibnamefont {Yan}},\ }\bibfield  {title} {\bibinfo {title} {Dynamic signatures of non-hermitian skin effect and topology in ultracold atoms},\ }\href {https://doi.org/10.1103/PhysRevLett.129.070401} {\bibfield  {journal} {\bibinfo  {journal} {Phys. Rev. Lett.}\ }\textbf {\bibinfo {volume} {129}},\ \bibinfo {pages} {070401} (\bibinfo {year} {2022})}\BibitemShut {NoStop}%
\bibitem [{\citenamefont {Guo}\ \emph {et~al.}(2022)\citenamefont {Guo}, \citenamefont {Dong}, \citenamefont {Zhang}, \citenamefont {Hu},\ and\ \citenamefont {Yang}}]{Guo2022}%
  \BibitemOpen
  \bibfield  {author} {\bibinfo {author} {\bibfnamefont {S.}~\bibnamefont {Guo}}, \bibinfo {author} {\bibfnamefont {C.}~\bibnamefont {Dong}}, \bibinfo {author} {\bibfnamefont {F.}~\bibnamefont {Zhang}}, \bibinfo {author} {\bibfnamefont {J.}~\bibnamefont {Hu}},\ and\ \bibinfo {author} {\bibfnamefont {Z.}~\bibnamefont {Yang}},\ }\bibfield  {title} {\bibinfo {title} {Theoretical prediction of a non-hermitian skin effect in ultracold-atom systems},\ }\href {https://doi.org/10.1103/PhysRevA.106.L061302} {\bibfield  {journal} {\bibinfo  {journal} {Phys. Rev. A}\ }\textbf {\bibinfo {volume} {106}},\ \bibinfo {pages} {L061302} (\bibinfo {year} {2022})}\BibitemShut {NoStop}%
\end{thebibliography}%


\begin{thebibliography}{16}%
\makeatletter
\providecommand \@ifxundefined [1]{%
 \@ifx{#1\undefined}
}%
\providecommand \@ifnum [1]{%
 \ifnum #1\expandafter \@firstoftwo
 \else \expandafter \@secondoftwo
 \fi
}%
\providecommand \@ifx [1]{%
 \ifx #1\expandafter \@firstoftwo
 \else \expandafter \@secondoftwo
 \fi
}%
\providecommand \natexlab [1]{#1}%
\providecommand \enquote  [1]{``#1''}%
\providecommand \bibnamefont  [1]{#1}%
\providecommand \bibfnamefont [1]{#1}%
\providecommand \citenamefont [1]{#1}%
\providecommand \href@noop [0]{\@secondoftwo}%
\providecommand \href [0]{\begingroup \@sanitize@url \@href}%
\providecommand \@href[1]{\@@startlink{#1}\@@href}%
\providecommand \@@href[1]{\endgroup#1\@@endlink}%
\providecommand \@sanitize@url [0]{\catcode `\\12\catcode `\$12\catcode `\&12\catcode `\#12\catcode `\^12\catcode `\_12\catcode `\%12\relax}%
\providecommand \@@startlink[1]{}%
\providecommand \@@endlink[0]{}%
\providecommand \url  [0]{\begingroup\@sanitize@url \@url }%
\providecommand \@url [1]{\endgroup\@href {#1}{\urlprefix }}%
\providecommand \urlprefix  [0]{URL }%
\providecommand \Eprint [0]{\href }%
\providecommand \doibase [0]{https://doi.org/}%
\providecommand \selectlanguage [0]{\@gobble}%
\providecommand \bibinfo  [0]{\@secondoftwo}%
\providecommand \bibfield  [0]{\@secondoftwo}%
\providecommand \translation [1]{[#1]}%
\providecommand \BibitemOpen [0]{}%
\providecommand \bibitemStop [0]{}%
\providecommand \bibitemNoStop [0]{.\EOS\space}%
\providecommand \EOS [0]{\spacefactor3000\relax}%
\providecommand \BibitemShut  [1]{\csname bibitem#1\endcsname}%
\let\auto@bib@innerbib\@empty
\bibitem [{\citenamefont {Resta}(1998)}]{Resta1998}%
  \BibitemOpen
  \bibfield  {author} {\bibinfo {author} {\bibfnamefont {R.}~\bibnamefont {Resta}},\ }\bibfield  {title} {\bibinfo {title} {Quantum-mechanical position operator in extended systems},\ }\href {https://doi.org/10.1103/PhysRevLett.80.1800} {\bibfield  {journal} {\bibinfo  {journal} {Phys. Rev. Lett.}\ }\textbf {\bibinfo {volume} {80}},\ \bibinfo {pages} {1800} (\bibinfo {year} {1998})}\BibitemShut {NoStop}%
\bibitem [{\citenamefont {Lee}\ \emph {et~al.}(2020)\citenamefont {Lee}, \citenamefont {Li}, \citenamefont {Thomale},\ and\ \citenamefont {Gong}}]{Lee2020}%
  \BibitemOpen
  \bibfield  {author} {\bibinfo {author} {\bibfnamefont {C.~H.}\ \bibnamefont {Lee}}, \bibinfo {author} {\bibfnamefont {L.}~\bibnamefont {Li}}, \bibinfo {author} {\bibfnamefont {R.}~\bibnamefont {Thomale}},\ and\ \bibinfo {author} {\bibfnamefont {J.}~\bibnamefont {Gong}},\ }\bibfield  {title} {\bibinfo {title} {Unraveling non-{H}ermitian pumping: Emergent spectral singularities and anomalous responses},\ }\href {https://doi.org/10.1103/PhysRevB.102.085151} {\bibfield  {journal} {\bibinfo  {journal} {Phys. Rev. B}\ }\textbf {\bibinfo {volume} {102}},\ \bibinfo {pages} {085151} (\bibinfo {year} {2020})}\BibitemShut {NoStop}%
\bibitem [{\citenamefont {L\"owdin}(1955)}]{Loewdin1955}%
  \BibitemOpen
  \bibfield  {author} {\bibinfo {author} {\bibfnamefont {P.-O.}\ \bibnamefont {L\"owdin}},\ }\bibfield  {title} {\bibinfo {title} {Quantum theory of many-particle systems. i. physical interpretations by means of density matrices, natural spin-orbitals, and convergence problems in the method of configurational interaction},\ }\href {https://doi.org/10.1103/PhysRev.97.1474} {\bibfield  {journal} {\bibinfo  {journal} {Phys. Rev.}\ }\textbf {\bibinfo {volume} {97}},\ \bibinfo {pages} {1474} (\bibinfo {year} {1955})}\BibitemShut {NoStop}%
\bibitem [{\citenamefont {Masuda}\ and\ \citenamefont {Nakamura}(2022)}]{Masuda2022a}%
  \BibitemOpen
  \bibfield  {author} {\bibinfo {author} {\bibfnamefont {S.}~\bibnamefont {Masuda}}\ and\ \bibinfo {author} {\bibfnamefont {M.}~\bibnamefont {Nakamura}},\ }\bibfield  {title} {\bibinfo {title} {Relationship between the electronic polarization and the winding number in non-{H}ermitian systems},\ }\href {https://doi.org/10.7566/JPSJ.91.043701} {\bibfield  {journal} {\bibinfo  {journal} {J. Phys. Soc. Jpn.}\ }\textbf {\bibinfo {volume} {91}},\ \bibinfo {pages} {043701} (\bibinfo {year} {2022})}\BibitemShut {NoStop}%
\bibitem [{\citenamefont {Ortega-Taberner}\ \emph {et~al.}(2022)\citenamefont {Ortega-Taberner}, \citenamefont {R\o{}dland},\ and\ \citenamefont {Hermanns}}]{OrtegaTaberner2022}%
  \BibitemOpen
  \bibfield  {author} {\bibinfo {author} {\bibfnamefont {C.}~\bibnamefont {Ortega-Taberner}}, \bibinfo {author} {\bibfnamefont {L.}~\bibnamefont {R\o{}dland}},\ and\ \bibinfo {author} {\bibfnamefont {M.}~\bibnamefont {Hermanns}},\ }\bibfield  {title} {\bibinfo {title} {Polarization and entanglement spectrum in non-{H}ermitian systems},\ }\href {https://doi.org/10.1103/PhysRevB.105.075103} {\bibfield  {journal} {\bibinfo  {journal} {Phys. Rev. B}\ }\textbf {\bibinfo {volume} {105}},\ \bibinfo {pages} {075103} (\bibinfo {year} {2022})}\BibitemShut {NoStop}%
\bibitem [{\citenamefont {Watanabe}\ and\ \citenamefont {Oshikawa}(2018)}]{Watanabe2018}%
  \BibitemOpen
  \bibfield  {author} {\bibinfo {author} {\bibfnamefont {H.}~\bibnamefont {Watanabe}}\ and\ \bibinfo {author} {\bibfnamefont {M.}~\bibnamefont {Oshikawa}},\ }\bibfield  {title} {\bibinfo {title} {Inequivalent berry phases for the bulk polarization},\ }\href {https://doi.org/10.1103/PhysRevX.8.021065} {\bibfield  {journal} {\bibinfo  {journal} {Phys. Rev. X}\ }\textbf {\bibinfo {volume} {8}},\ \bibinfo {pages} {021065} (\bibinfo {year} {2018})}\BibitemShut {NoStop}%
\bibitem [{\citenamefont {Zaletel}\ \emph {et~al.}(2014)\citenamefont {Zaletel}, \citenamefont {Mong},\ and\ \citenamefont {Pollmann}}]{Zaletel2014}%
  \BibitemOpen
  \bibfield  {author} {\bibinfo {author} {\bibfnamefont {M.~P.}\ \bibnamefont {Zaletel}}, \bibinfo {author} {\bibfnamefont {R.~S.~K.}\ \bibnamefont {Mong}},\ and\ \bibinfo {author} {\bibfnamefont {F.}~\bibnamefont {Pollmann}},\ }\bibfield  {title} {\bibinfo {title} {Flux insertion, entanglement, and quantized responses},\ }\href {https://doi.org/10.1088/1742-5468/2014/10/P10007} {\bibfield  {journal} {\bibinfo  {journal} {J. Stat. Mech. Theor. Exp.}\ }\textbf {\bibinfo {volume} {2014}},\ \bibinfo {pages} {P10007} (\bibinfo {year} {2014})}\BibitemShut {NoStop}%
\bibitem [{\citenamefont {Kohn}(1996)}]{Kohn1996}%
  \BibitemOpen
  \bibfield  {author} {\bibinfo {author} {\bibfnamefont {W.}~\bibnamefont {Kohn}},\ }\bibfield  {title} {\bibinfo {title} {Density functional and density matrix method scaling linearly with the number of atoms},\ }\href {https://doi.org/10.1103/PhysRevLett.76.3168} {\bibfield  {journal} {\bibinfo  {journal} {Phys. Rev. Lett.}\ }\textbf {\bibinfo {volume} {76}},\ \bibinfo {pages} {3168} (\bibinfo {year} {1996})}\BibitemShut {NoStop}%
\bibitem [{\citenamefont {King-Smith}\ and\ \citenamefont {Vanderbilt}(1993)}]{KingSmith1993}%
  \BibitemOpen
  \bibfield  {author} {\bibinfo {author} {\bibfnamefont {R.~D.}\ \bibnamefont {King-Smith}}\ and\ \bibinfo {author} {\bibfnamefont {D.}~\bibnamefont {Vanderbilt}},\ }\bibfield  {title} {\bibinfo {title} {Theory of polarization of crystalline solids},\ }\href {https://doi.org/10.1103/PhysRevB.47.1651} {\bibfield  {journal} {\bibinfo  {journal} {Phys. Rev. B}\ }\textbf {\bibinfo {volume} {47}},\ \bibinfo {pages} {1651} (\bibinfo {year} {1993})}\BibitemShut {NoStop}%
\bibitem [{\citenamefont {Peschel}(2003)}]{Peschel2003}%
  \BibitemOpen
  \bibfield  {author} {\bibinfo {author} {\bibfnamefont {I.}~\bibnamefont {Peschel}},\ }\bibfield  {title} {\bibinfo {title} {Calculation of reduced density matrices from correlation functions},\ }\href {https://doi.org/10.1088/0305-4470/36/14/101} {\bibfield  {journal} {\bibinfo  {journal} {J. Phys. A: Math. Theor.}\ }\textbf {\bibinfo {volume} {36}},\ \bibinfo {pages} {L205} (\bibinfo {year} {2003})}\BibitemShut {NoStop}%
\bibitem [{\citenamefont {A}\ and\ \citenamefont {B}(2006)}]{Bottcher2006}%
  \BibitemOpen
  \bibfield  {author} {\bibinfo {author} {\bibfnamefont {B.}~\bibnamefont {A}}\ and\ \bibinfo {author} {\bibfnamefont {S.}~\bibnamefont {B}},\ }\href@noop {} {\emph {\bibinfo {title} {Analysis of Toeplitz operators}}}\ (\bibinfo  {publisher} {Berlin, Heidelberg: Springer Berlin Heidelberg},\ \bibinfo {year} {2006})\BibitemShut {NoStop}%
\bibitem [{\citenamefont {Ortega-Taberner}\ and\ \citenamefont {Hermanns}(2021)}]{OrtegaTaberner2021}%
  \BibitemOpen
  \bibfield  {author} {\bibinfo {author} {\bibfnamefont {C.}~\bibnamefont {Ortega-Taberner}}\ and\ \bibinfo {author} {\bibfnamefont {M.}~\bibnamefont {Hermanns}},\ }\bibfield  {title} {\bibinfo {title} {Relation of the entanglement spectrum to the bulk polarization},\ }\href {https://doi.org/10.1103/PhysRevB.103.195132} {\bibfield  {journal} {\bibinfo  {journal} {Phys. Rev. B}\ }\textbf {\bibinfo {volume} {103}},\ \bibinfo {pages} {195132} (\bibinfo {year} {2021})}\BibitemShut {NoStop}%
\bibitem [{\citenamefont {Yao}\ and\ \citenamefont {Wang}(2018)}]{Yao2018}%
  \BibitemOpen
  \bibfield  {author} {\bibinfo {author} {\bibfnamefont {S.}~\bibnamefont {Yao}}\ and\ \bibinfo {author} {\bibfnamefont {Z.}~\bibnamefont {Wang}},\ }\bibfield  {title} {\bibinfo {title} {Edge states and topological invariants of non-hermitian systems},\ }\href {https://doi.org/10.1103/PhysRevLett.121.086803} {\bibfield  {journal} {\bibinfo  {journal} {Phys. Rev. Lett.}\ }\textbf {\bibinfo {volume} {121}},\ \bibinfo {pages} {086803} (\bibinfo {year} {2018})}\BibitemShut {NoStop}%
\bibitem [{\citenamefont {Yokomizo}\ and\ \citenamefont {Murakami}(2019)}]{Yokomizo2019}%
  \BibitemOpen
  \bibfield  {author} {\bibinfo {author} {\bibfnamefont {K.}~\bibnamefont {Yokomizo}}\ and\ \bibinfo {author} {\bibfnamefont {S.}~\bibnamefont {Murakami}},\ }\bibfield  {title} {\bibinfo {title} {Non-{B}loch band theory of non-{H}ermitian systems},\ }\href {https://doi.org/10.1103/PhysRevLett.123.066404} {\bibfield  {journal} {\bibinfo  {journal} {Phys. Rev. Lett.}\ }\textbf {\bibinfo {volume} {123}},\ \bibinfo {pages} {066404} (\bibinfo {year} {2019})}\BibitemShut {NoStop}%
\bibitem [{\citenamefont {Hatano}\ and\ \citenamefont {Nelson}(1996)}]{Hatano1996}%
  \BibitemOpen
  \bibfield  {author} {\bibinfo {author} {\bibfnamefont {N.}~\bibnamefont {Hatano}}\ and\ \bibinfo {author} {\bibfnamefont {D.~R.}\ \bibnamefont {Nelson}},\ }\bibfield  {title} {\bibinfo {title} {Localization transitions in non-hermitian quantum mechanics},\ }\href {https://doi.org/10.1103/PhysRevLett.77.570} {\bibfield  {journal} {\bibinfo  {journal} {Phys. Rev. Lett.}\ }\textbf {\bibinfo {volume} {77}},\ \bibinfo {pages} {570} (\bibinfo {year} {1996})}\BibitemShut {NoStop}%
\bibitem [{\citenamefont {Lee}(2022)}]{Lee2022a}%
  \BibitemOpen
  \bibfield  {author} {\bibinfo {author} {\bibfnamefont {C.~H.}\ \bibnamefont {Lee}},\ }\bibfield  {title} {\bibinfo {title} {Exceptional bound states and negative entanglement entropy},\ }\href {https://doi.org/10.1103/PhysRevLett.128.010402} {\bibfield  {journal} {\bibinfo  {journal} {Phys. Rev. Lett.}\ }\textbf {\bibinfo {volume} {128}},\ \bibinfo {pages} {010402} (\bibinfo {year} {2022})}\BibitemShut {NoStop}%
\end{thebibliography}%

\end{document}


\onecolumngrid
\newpage
\renewcommand{\theequation}{S\arabic{equation}}
\renewcommand{\thefigure}{S\arabic{figure}}
\renewcommand{\thetable}{S\arabic{table}}
\setcounter{equation}{0}
\setcounter{figure}{0}
\setcounter{table}{0}

\begin{center}
\textbf{\large Supplementary Material for "Unified Bulk–Entanglement Correspondence in Non-Hermitian Systems"}
\label{sec: SM}
\end{center}

\tableofcontents

\section{S1. Equivalence of Real-Space Resta and Non-Bloch Momentum-Space Polarizations}
\label{sec: S1}

This derivation is performed in the quasi-local limit, where the generalized Brillouin zone (GBZ) contour $C_\beta$ forms a perfect circle $\beta(k) = r_0 e^{ik}$ with a constant radius $r_0$. The quasi-reciprocal Hamiltonian $\tilde{H}$ is obtained via the inverse Fourier transform of $H(\beta)$ along this circular $C_\beta$. Its momentum-space Hamiltonian $H_{\tilde{H}}(k)$—the standard Bloch transform of $\tilde{H}$—is by definition equal to $H(\beta)$ evaluated on $C_\beta$, i.e., $H_{\tilde{H}}(k) = H(\beta(k)) = H(r_0 e^{ik})$.

Since $\tilde{H}$ is a quasi-local periodic Hamiltonian in this limit (with exponentially decaying hoppings), its eigenstates under periodic boundary conditions (PBC) strictly obey the standard Bloch theorem. The eigenstates take the explicit form $\psi_{kR}(x) = e^{ikx} |u_{\beta(k)R}\rangle$ and $\psi_{kL}(x) = e^{ikx} |u_{\beta(k)L}\rangle$, where $k$ denotes the real momentum in the standard Brillouin zone, and $|u_{\beta(k)}\rangle$ represents the cell-periodic eigenstate of $H(\beta)$ at $\beta=\beta(k)$. This expression establishes the crucial connection between the real-space lattice model $\tilde{H}$ and the auxiliary GBZ Hamiltonian $H(\beta)$.

We begin with the Resta polarization $P_{\tilde{H}}$ for $\tilde{H}$. Its value is determined by the expectation value of the many-body position operator twist $e^{i\Delta k \hat{X}}$ (with $\Delta k = 2\pi/L$) in the many-body ground state $|\Psi\rangle$~\cite{Resta1998, Lee2020}:
\begin{equation}
    e^{i2\pi P_{\tilde{H}}} = \langle\Psi^{L}|e^{i\Delta k \hat{X}}|\Psi^{R}\rangle,
\end{equation}
where $|\Psi^R\rangle$ and $\langle\Psi^L|$ are the Slater determinants constructed from the $N$ occupied single-particle wavefunctions $\psi_{kR}$ and $\psi_{kL}$, respectively.

By the generalized L\"{o}wdin rule~\cite{Loewdin1955}, this many-body overlap reduces to the determinant of the $N \times N$ single-particle overlap matrix $S$:
\begin{equation}
    e^{i2\pi P_{\tilde{H}}} = \det[S],
\end{equation}
where the matrix elements are given by $S_{k, k'} = \langle \psi_{kL} | e^{i\Delta k \hat{X}} | \psi_{k'R} \rangle$.

Substituting the explicit Bloch wavefunction forms into the matrix element yields:
\begin{align}
    S_{k, k'} &= \int_0^L dx \, (e^{ikx} u_{\beta(k)L}(x))^* \, (e^{i\Delta k x} [e^{ik'x} u_{\beta(k')R}(x)]) \nonumber\\
    &= \int_0^L dx \, e^{i(k' - k + \Delta k)x} [u_{\beta(k)L}^*(x) u_{\beta(k')R}(x)].
\end{align}

Since both $k$ and $k'$ are real momenta quantized in the standard Brillouin zone, the bracketed term is cell-periodic. By the orthogonality condition of the standard Bloch theorem, the integral over the phase factor vanishes unless the total momentum transfer is zero, i.e., $k' - k + \Delta k = 0$. This enforces the selection rule:
\begin{equation}
    k' = k - \Delta k.
\end{equation}

This sparsity condition implies that $S$ is a cyclic shift matrix. Its determinant is simply the product of all non-zero off-diagonal elements, multiplied by a sign factor arising from the permutation parity:
\begin{align}
    e^{i2\pi P_{\tilde{H}}} = \det[S] = (-1)^{\frac{L-1}{2}M} \prod_{k \in \text{occ}} S_{k, k-\Delta k},
\end{align}
where $M$ denotes the number of occupied bands ($M=1$ for single-band models).

The non-zero matrix element $S_{k, k-\Delta k}$ evaluates to the biorthogonal overlap of the cell-periodic parts:
\begin{align}
    S_{k, k-\Delta k} &= \int_0^L dx \, e^{i((k-\Delta k) - k + \Delta k)x} [u_{\beta(k)L}^*(x) u_{\beta(k-\Delta k)R}(x)] \nonumber\\
    &= \int_0^L dx \, u_{\beta(k)L}^*(x) u_{\beta(k-\Delta k)R}(x) = \langle u_{\beta(k)L} | u_{\beta(k - \Delta k)R} \rangle.
\end{align}

Substituting this back into the determinant expression yields:
\begin{align}\label{Eq: S6}
    e^{i2\pi P_{\tilde{H}}} = (-1)^{\frac{L-1}{2}M} \prod_{k \in \text{occ}} \langle u_{\beta(k)L} | u_{\beta(k - \Delta k)R} \rangle.
\end{align}

The product term is precisely the discretized non-Bloch Wilson loop $W_\beta$ defined along the GBZ contour $C_\beta$:
\begin{align} \label{Eq: S7}
    e^{i2\pi P_{\tilde{H}}} = (-1)^{\frac{L-1}{2}M} W_\beta.
\end{align}

The prefactor $(-1)^{\frac{L-1}{2}M}$ depends only on the system size $L$ and band number $M$, representing a trivial topological offset of $0$ or $1/2$. In the thermodynamic limit ($L \to \infty$), the phase of the Wilson loop defines the non-Bloch Zak phase $\gamma_\beta$, such that $W_\beta \to e^{i\gamma_\beta}$. The non-Bloch polarization $P_\beta$ is defined as this winding number normalized by $2\pi$~\cite{Masuda2022a, OrtegaTaberner2022}:
\begin{align}
    P_\beta = \frac{\gamma_\beta}{2\pi} = \frac{1}{2\pi i} \ln(W_\beta) \equiv \frac{1}{2\pi i} \oint_{C_\beta} \mathcal{A}(\beta) d\beta \pmod 1.
\end{align}

Thus, we have rigorously demonstrated that in the quasi-local limit, the real-space Resta polarization of $\tilde{H}$ is equivalent to the momentum-space non-Bloch polarization:
\begin{align}
    P_{\text{Resta}}(\tilde{H}) \equiv P_\beta \pmod 1.
\end{align}

It is crucial to note that this derivation relies on the standard Bloch theorem ($\psi_k = e^{ikx} u_k$), which holds only when $\tilde{H}$ is a local (or quasi-local) Hamiltonian. In the general non-local regime (e.g., $t_3 \neq 0$), the GBZ becomes non-circular, making $\beta(k)$ a non-analytic function of real momentum $k$. This non-analyticity generates power-law long-range hoppings in $\tilde{H}$, causing the position variance $\langle \hat{X}^2 \rangle$ to diverge (violation of the nearsightedness principle). Consequently, the momentum selection rule $k' = k - \Delta k$ breaks down, the overlap matrix $S$ loses its sparsity, and the equivalence in Eq.~(\ref{Eq: S6}) is invalidated. This explains the failure of $P_{\text{Resta}}$ observed in the main text.

To bridge the gap to the entanglement polarization, we introduce the Biorthogonal Flux Insertion Polarization $P_{\text{Flux}}$. This quantity is defined as the Berry phase accumulated by the many-body ground state $|\tilde{\Phi}(\theta)\rangle$ during the adiabatic insertion of a $2\pi$ magnetic flux $\theta$~\cite{Watanabe2018, OrtegaTaberner2022, Zaletel2014}:
\begin{align}\label{Eq: S12}
    P_{\text{Flux}} = i \int_0^{2\pi} \frac{d\theta}{2\pi} \langle\tilde{\Phi}^L(\theta)|\partial_{\theta}|\tilde{\Phi}^R(\theta)\rangle.
\end{align}

Following the decomposition analysis in Ref.~\cite{OrtegaTaberner2022}, this many-body phase splits into a boundary term and a bulk term. The bulk contribution, after the substitution $d\theta/L \to dk$, maps directly to the integral of the Berry connection over the Brillouin zone, which in our case corresponds to the non-Bloch Zak phase $\gamma_\beta$:
\begin{align}
    P_{\text{Flux}} \equiv \frac{1}{2\pi} \int_0^{2\pi} dk \sum_{\mu} \langle u_{\beta(k)\mu L} | i\partial_k | u_{\beta(k)\mu R} \rangle = \frac{\gamma_\beta}{2\pi} \equiv P_\beta \pmod 1.
\end{align}

We have thus established the chain of equivalences in the quasi-local limit:
\begin{align}
    P_{\text{Resta}}(\tilde{H})\equiv P_{\text{Flux}}(\tilde{H}) \equiv P_\beta.
\end{align}

This $P_{\text{Flux}}$ serves as the crucial intermediate link. In Section~\hyperref[sec: S2]{S2}, we will show that $P_{\text{Flux}}$ is identically equal to the entanglement polarization $\chi(\tilde{H})$, thereby completing the proof.

\section{S2. Equivalence Between Flux Insertion Polarization and Entanglement Spectrum Polarization}
\label{sec: S2}

Having established the equivalence $P_{\text{Resta}}(\tilde{H}) \equiv P_{\text{Flux}}(\tilde{H}) \equiv P_\beta$ in Section~\hyperref[sec: S1]{S1}, we now demonstrate the final link: the equivalence between the many-body flux polarization $P_{\text{Flux}}$ and the single-particle entanglement polarization $\chi(\tilde{H})$. This completes the proof that $P_\beta \equiv \chi(\tilde{H})$.

We employ the formulation of $P_{\text{Flux}}$ based on the twist operator method~\cite{Kohn1996, KingSmith1993, OrtegaTaberner2022}. Inserting a magnetic flux $\theta$ at the cut between subsystems $A$ and $B$ is mathematically equivalent to applying a large gauge transformation $U(\theta) = e^{-i\theta \hat{N}_A}$ restricted to subsystem $A$, where $\hat{N}_A = \sum_{j \in A} c_j^\dagger c_j$ is the particle number operator in region $A$. The definition of $P_{\text{Flux}}$ (from Eq.~(\ref{Eq: S12}) in Section~\hyperref[sec: S1]{S1}) can be rewritten as:
\begin{align}
    P_{\text{Flux}} = i \int_0^{2\pi} \frac{d\theta}{2\pi} \langle\tilde{\Psi}^L| U^\dagger(\theta) (\partial_{\theta} U(\theta)) |\tilde{\Psi}^R\rangle.
\end{align}

Substituting the explicit form of the unitary operator $\partial_\theta U(\theta) = -i \hat{N}_A e^{-i\theta \hat{N}_A}$ and noting that $U^\dagger(\theta)$ commutes with $\hat{N}_A$, we obtain:
\begin{align}
    P_{\text{Flux}} &= i \int_0^{2\pi} \frac{d\theta}{2\pi} \langle\tilde{\Psi}^L| e^{i\theta \hat{N}_A} (-i \hat{N}_A) e^{-i\theta \hat{N}_A} |\tilde{\Psi}^R\rangle \nonumber \\
    &= \int_0^{2\pi} \frac{d\theta}{2\pi} \langle\tilde{\Psi}^L| \hat{N}_A |\tilde{\Psi}^R\rangle.
\end{align}

The integrand is independent of $\theta$, so the integration yields unity, leading to the identification of flux polarization as the fractional boundary charge:
\begin{align}\label{eq:PfluxEqNA}
    P_{\text{Flux}}(\tilde{H}) = \langle\tilde{\Psi}^L| \hat{N}_A |\tilde{\Psi}^R\rangle \pmod 1.
\end{align}

Next, we connect this many-body expectation value to the entanglement spectrum. We define the biorthogonal reduced density matrix of subsystem $A$ as $\rho_A = \mathrm{Tr}_B(|\tilde{\Psi}^R\rangle\langle\tilde{\Psi}^L|)$. The expectation value of the particle number in subsystem $A$ is given by:
\begin{align}
    \langle \hat{N}_A \rangle = \mathrm{Tr}_{A+B}\left( |\tilde{\Psi}^R\rangle\langle\tilde{\Psi}^L| \hat{N}_A \right).
\end{align}

By the definition of the partial trace and the cyclic property, this reduces to:
\begin{align}
    \mathrm{Tr}_{A+B}\left( |\tilde{\Psi}^R\rangle\langle\tilde{\Psi}^L| \hat{N}_A \right) = \mathrm{Tr}_A\left( \mathrm{Tr}_B(|\tilde{\Psi}^R\rangle\langle\tilde{\Psi}^L|) \hat{N}_A \right) = \mathrm{Tr}_A(\rho_A \hat{N}_A).
\end{align}

Combining this with Eq.~(\ref{eq:PfluxEqNA}), we verify that the flux polarization is determined by the reduced density matrix:
\begin{align}
    P_{\text{Flux}}(\tilde{H}) = \mathrm{Tr}_A(\rho_A \hat{N}_A).
\end{align}

Finally, we invoke the property that for non-interacting fermions, the many-body reduced density matrix $\rho_A$ is Gaussian and completely characterized by the single-particle correlation matrix $C^A_{ij} = \langle \tilde{\Psi}^L | c^\dagger_i c_j | \tilde{\Psi}^R \rangle$ (where $i,j \in A$). This constitutes the biorthogonal generalization of the Peschel-type relation~\cite{OrtegaTaberner2022, Peschel2003}. Specifically, the trace of the particle number operator relates directly to the eigenvalues $\{\xi_\mu\}$ of the correlation matrix $C^A$:
\begin{align}
    \mathrm{Tr}_A(\rho_A \hat{N}_A) = \mathrm{Tr}(C^A) = \sum_{\mu \in L} \xi_\mu \pmod 1.
\end{align}
Note that the modulo 1 ambiguity arises because the integer part of the charge corresponds to filled bulk bands, while the fractional part corresponds to the polarization $\chi(\tilde{H})$ defined in Eq.~(\ref{Eq.4}) of the main text.

Thus, we have established the identity:
\begin{align}
    P_{\text{Flux}}(\tilde{H}) \equiv \chi(\tilde{H}) \pmod 1.
\end{align}

Combining the results from Section~\hyperref[sec: S1]{S1} and Section~\hyperref[sec: S2]{S2}, we have rigorously proven the following chain of equivalences in the quasi-local limit:
\begin{align} \label{eq:chain}
    P_{\text{Resta}}(\tilde{H}) \equiv P_{\text{Flux}}(\tilde{H}) \equiv P_\beta \equiv \chi(\tilde{H}) \pmod 1.
\end{align}

This analytical proof serves as the baseline for our theory. Crucially, while the first link in this chain ($P_{\text{Resta}}$) relies on the position operator and breaks down in the non-local regime due to variance divergence, the final link ($\chi(\tilde{H})$) relies on the algebraic trace properties of the correlation matrix. In Section~\hyperref[sec: S3]{S3}, we will demonstrate that this algebraic definition $\chi(\tilde{H})$ remains robust even when locality is lost, protected by the index theory of Toeplitz operators.

\section{S3. Theoretical Robustness of Entanglement Polarization in the Non-Local Regime}
\label{sec: S3}

While the equivalence $P_\beta \equiv \chi(\tilde{H})$ was rigorously derived in Sections~\hyperref[sec: S1]{S1} and~\hyperref[sec: S2]{S2} under the quasi-locality assumption (exponential decay of hoppings), our numerical results demonstrate that this correspondence persists even when $\tilde{H}$ becomes effectively non-local due to a non-circular GBZ. In this regime, the effective hoppings of $\tilde{H}$ decay as a power law, $\tilde{t}_n \sim |n|^{-\alpha}$. This leads to a crucial theoretical distinction between the fragility of the geometric Resta polarization and the robustness of the algebraic entanglement polarization.

\subsection{A. Geometric Fragility vs. Algebraic Robustness}

The breakdown of the Resta polarization $P_{\text{Resta}}(\tilde{H})$ in the non-local regime originates from its definition via the position operator twist $e^{i \Delta k \hat{X}}$. As discussed in S1, for this quantity to be well-defined and quantized, the ground state $|\Psi\rangle$ must satisfy the ``nearsightedness'' principle, which requires the variance of the position operator, $\sigma_X^2 = \langle \hat{X}^2 \rangle - \langle \hat{X} \rangle^2$, to be finite.

In the presence of power-law hoppings $\tilde{t}_n \sim |n|^{-\alpha}$, the Wannier functions develop heavy tails. Specifically, for one-dimensional systems, it is known that if the decay rate satisfies $\alpha \le 1.5$, the position variance $\sigma_X^2$ diverges~\cite{Resta1998, Kohn1996}. This divergence introduces extreme phase fluctuations in the evaluation of $\langle \Psi | e^{i \frac{2\pi}{L} \hat{X}} | \Psi \rangle$, rendering $P_{\text{Resta}}(\tilde{H})$ numerically unstable and physically ill-defined.

In stark contrast, the entanglement polarization $\chi(\tilde{H})$ is derived solely from the spectrum of the biorthogonal correlation matrix $C_{ij} = \langle \tilde{\Psi}^L | c^\dagger_i c_j | \tilde{\Psi}^R \rangle$. Its stability relies not on the spatial localization moments (geometry), but on the existence of a spectral gap in the correlation matrix (algebra). As long as the entanglement gap remains open, $\chi(\tilde{H})$ remains a robust topological invariant.

\subsection{B. Protection via Toeplitz Index Theory}

The robustness of $\chi(\tilde{H})$ against non-locality can be rigorously grounded in the index theory of Toeplitz operators~\cite{Bottcher2006}. In the thermodynamic limit ($L \to \infty$), the biorthogonal correlation matrix $C$ converges to a Toeplitz matrix generated by a "Symbol" function $\Xi(\beta)$. Physically, this Symbol corresponds to the projection operator in the GBZ space.

According to the Widom-Dyson theorem (and its non-Hermitian generalizations)~\cite{ Bottcher2006}, the Fredholm index of a Toeplitz operator is well-defined and quantized if its Symbol is continuous and non-vanishing on the integration contour. In real space, the continuity of the Symbol requires the Fourier coefficients (the effective hoppings $\tilde{t}_n$) to be absolutely summable or to decay sufficiently fast. Specifically, the condition for the Symbol to belong to the Wiener algebra (ensuring boundedness and Fredholm property) is~\cite{Bottcher2006}
\begin{equation}
    \alpha > 1.
\end{equation}

This establishes a critical window of robustness:
\begin{itemize}
    \item When $\alpha \le 1.5$, the geometric Resta polarization fails due to the divergence of $\langle \hat{X}^2 \rangle$.
    \item However, as long as $\alpha > 1$, the algebraic entanglement polarization $\chi(\tilde{H})$ remains topologically protected by the Toeplitz index.
\end{itemize}

For the non-Hermitian models considered in this work, the deformation of the GBZ arises from algebraic characteristic equations. Consequently, the singularities of the Symbol are algebraic branch points rather than essential singularities. This algebraic nature typically imposes a decay rate (e.g., $\alpha=1.5$ for square-root branch points) that satisfies $\alpha \ge 1$, thereby falling precisely into the regime where Resta fails but $\chi$ survives.

\subsection{C. Symmetry Protection}

Finally, we confirm that this topological index maps precisely to $0$ or $0.5$ due to chiral symmetry $\Gamma$. Since the surrogate Hamiltonian $\tilde{H}$ is derived from the chiral-symmetric $H(\beta)$ via a linear inverse Fourier transform, $\tilde{H}$ strictly preserves the symmetry: $\Gamma \tilde{H} \Gamma^{-1} = -\tilde{H}$.

This symmetry extends to the correlation matrix, implying $\Gamma (C - 1/2) \Gamma^{-1} = -(C - 1/2)$. Consequently, the entanglement spectrum $\{\xi_\mu\}$ must be symmetric about $\xi = 1/2$. Eigenvalues occur in pairs $(\xi, 1-\xi)$, which sum to $1$ and vanish modulo $1$. The only contribution to the polarization $\chi(\tilde{H}) = \sum \xi_\mu$ comes from unpaired self-conjugate modes pinned strictly at $\xi = 1/2$. Since these zero modes are protected by the algebraic symmetry $\Gamma$ and their number is counted by the robust Toeplitz Index, $\chi(\tilde{H})$ remains exactly quantized to $P_\beta$ even when real-space locality is compromised.

\section{S4. Robust Calculation of Polarization from Degenerate Entanglement Spectrum}
\label{sec: S4}

In non-Hermitian systems, the calculation of the entanglement polarization $\chi(\tilde{H})$ is often complicated by the presence of degeneracies in the entanglement spectrum. Standard methods that sum eigenvalues based on simple spatial sorting may fail when degenerate subspaces mix states from different boundaries. Here, we generalize the projection-based method~\cite{OrtegaTaberner2021} to the biorthogonal framework to resolve this issue.

Consider a $d$-fold degenerate entanglement eigenvalue $\xi_\lambda$. A numerical diagonalization routine will return a set of $d$ biorthonormal eigenvectors $\{|\psi_{\mu}^{R}\rangle, \langle\psi_{\mu}^{L}|\}_{\mu=1}^d$ spanning the degenerate subspace $\mathcal{H}_\lambda$. These vectors satisfy the biorthogonality condition $\langle\psi_{\mu}^{L}|\psi_{\nu}^{R}\rangle = \delta_{\mu\nu}$. Crucially, the numerical basis vectors are generally arbitrary linear combinations of the "physical" modes, which may be localized at opposite boundaries.

To extract the topological contribution, we must determine the integer number of states within this subspace that are physically localized at the left ($L$) boundary, denoted as $N_L(\lambda)$. We first construct the biorthogonal projector onto the degenerate subspace $\mathcal{H}_\lambda$:
\begin{align}
    P_{\lambda} = \sum_{\mu=1}^{d} |\psi_{\mu}^{R}\rangle\langle\psi_{\mu}^{L}|.
\end{align}
This operator is invariant under any linear basis transformation within the subspace.

Next, we define the spatial projector onto the left partition of the system:
\begin{equation}
    \Pi_{L} = \sum_{j \in L} |j\rangle\langle j| + \sum_{j \in L} |j'\rangle\langle j'|,
\end{equation}
where the sum runs over all degrees of freedom (unit cells and internal orbitals) within the left subsystem.

The number of left-localized modes is then given by the trace of the projected density:
\begin{equation}
    N_L(\lambda) = \mathrm{Tr}(P_\lambda \Pi_L).
\end{equation}

In the thermodynamic limit, an eigenstate is either fully localized in $L$ (eigenvalue of $\Pi_L \approx 1$) or fully localized in $R$ (eigenvalue of $\Pi_L \approx 0$). Thus, the trace converges to an integer:
\begin{align}
    N_L(\lambda) = \sum_{\mu=1}^d \langle \psi_\mu^L | \Pi_L | \psi_\mu^R \rangle
\end{align}
This formula is robust because it relies on the basis-independent trace operation. Even if the numerical solver returns a basis that mixes left and right modes, the trace operation correctly disentangles their spatial weights.

Finally, the total entanglement polarization is obtained by summing the eigenvalues weighted by their left-occupation integers:
\begin{equation}
    \chi(\tilde{H}) = \sum_{\lambda} N_L(\lambda) \,\xi_{\lambda} \pmod{1}.
\end{equation}

This method ensures that $\chi(\tilde{H})$ is calculated accurately even in the presence of mixing between degenerate edge modes or between edge and bulk bands, providing a numerically robust implementation of the real-space invariant.

\section{S5. The One-Dimensional Non-Hermitian Hatano-Nelson Type Model}
\label{sec: S5}

\begin{figure}[t]
\centering
\includegraphics[width=0.8\textwidth,keepaspectratio]{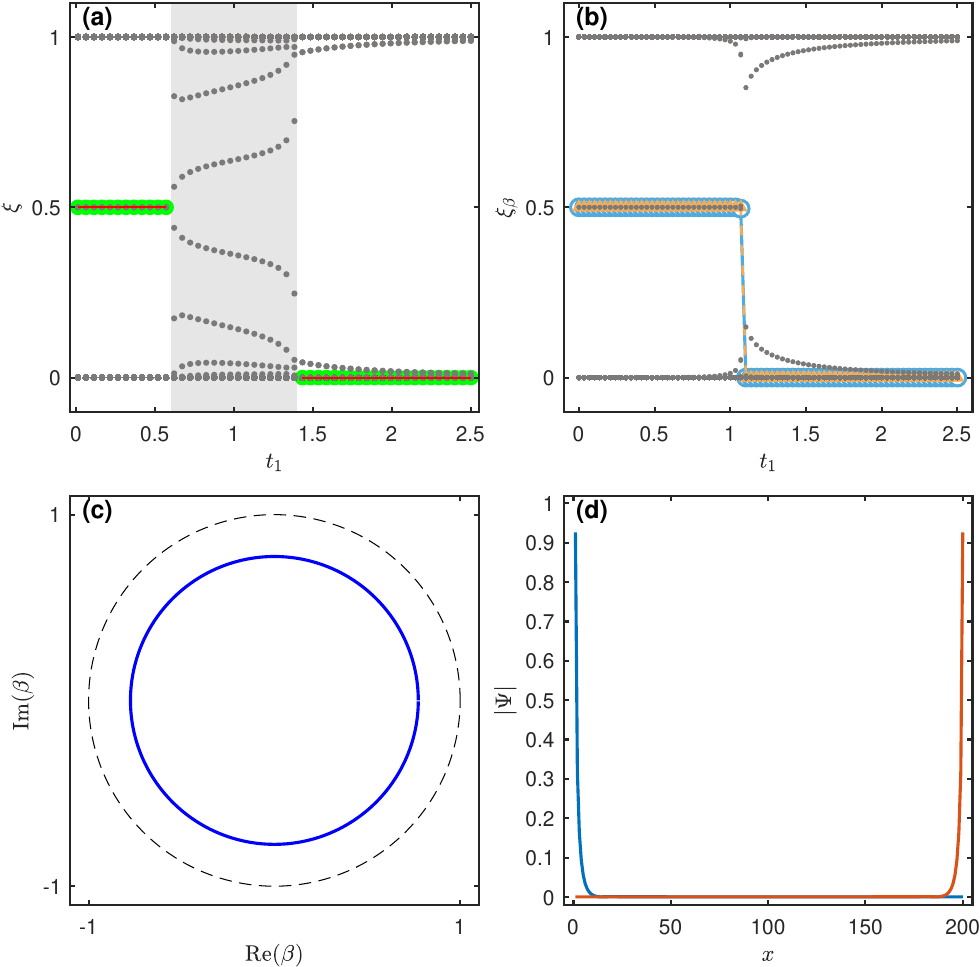}
\caption{Verification in the quasi-local limit ($t_3=0$). (a) PBC entanglement spectrum (grey) vs $t_1$. Green dots: PBC entanglement polarization $\chi$. Blue line: Bloch polarization $P$. Note the failure of PBC invariants in the point-gapped regime ($0.6 < t_1 < 1.4$). (b) Perfect agreement between $P_\beta$ (yellow triangles) and $\chi(\tilde{H})$ (blue circles), validating the theory in the circular-GBZ limit. (c) The circular GBZ (blue solid) for $t_1 = 0.8$, indicating quasi-local $\tilde{H}$. Grey dashed: unit circle. (d) Spatial localization of $\xi=0.5$ modes confirms the BBC. Parameters: $t_2 = 1$, $\gamma = 0.4$, $t_3 = 0$, $L = 800$.}
\label{Fig_S1}
\end{figure}

This section details the non-Hermitian model Hamiltonian and the formalism of non-Bloch band theory used to validate the core equivalence $P_\beta \equiv \chi(\tilde{H}) \pmod 1$. We employ a prototypical one-dimensional non-Hermitian Hatano-Nelson type model~\cite{Yao2018, Yokomizo2019, Hatano1996}. While this model serves as a concrete example, our theoretical framework is universally applicable to general 1D non-Hermitian free-fermion systems.

The model is defined on a 1D lattice with a two-site unit cell ($A$ and $B$ sublattices). The Hamiltonian $H$ includes intra-cell hopping ($t_1$ with non-reciprocity $\gamma$) and inter-cell hoppings ($t_2$ and $t_3$ for nearest- and next-nearest-neighbor couplings, respectively):
\begin{align}
    H = \sum_n \Big[ 
        (t_1 + \gamma)c^\dagger_{n,A}c_{n,B} + (t_1 - \gamma)c^\dagger_{n,B}c_{n,A} 
        + t_2 \left(c^\dagger_{n,B}c_{n+1,A} + c^\dagger_{n+1,A}c_{n,B}\right) 
        + t_3 \left(c^\dagger_{n,A}c_{n+1,B} + c^\dagger_{n+1,B}c_{n,A}\right)
    \Big].
\end{align}

Its Bloch Hamiltonian $H(k)$ is obtained via standard Fourier transformation:
\begin{align}
    H(k) = \begin{pmatrix} 0 & (t_1 + \gamma) + t_2e^{-ik} + t_3e^{ik} \\ (t_1 - \gamma) + t_3e^{-ik} + t_2e^{ik} & 0 \end{pmatrix},
\end{align}
where all parameters $t_1, t_2, t_3, \gamma$ are real. The non-reciprocal term ($\gamma \ne 0$) induces the Non-Hermitian Skin Effect (NHSE).

To properly treat the NHSE and restore the bulk-boundary correspondence (BBC), we adopt NBBT by analytically continuing the Bloch phase factor $e^{ik}$ to a general complex variable $\beta$. The non-Bloch Hamiltonian $H(\beta)$ is obtained by the substitution $e^{ik} \to \beta$:
\begin{align}
    H(\beta) = \begin{pmatrix} 0 & (t_1 + \gamma) + t_2\beta^{-1} + t_3\beta \\ (t_1 - \gamma) + t_3\beta^{-1} + t_2\beta & 0 \end{pmatrix} 
    \equiv \begin{pmatrix} 0 & R_+(\beta) \\ R_-(\beta) & 0 \end{pmatrix}.
\end{align}

The characteristic equation $\det[H(\beta) - E] = 0$ determines the GBZ contour $C_\beta$. According to NBBT, $C_\beta$ is the trajectory of $\beta$ satisfying the condition $|\beta_M(E)| = |\beta_{M+1}(E)|$ ($M=2$ for this model) for bulk energy bands~\cite{Yao2018, Yokomizo2019}.

The key parameter governing the locality of the quasi-reciprocal system $\tilde{H}$ is $t_3$. When $t_3 = 0$, the GBZ contour $C_{\beta}$ remains a simple circle ($\beta = r_0 e^{ik}$). This analytic smoothness ensures that the hoppings of $\tilde{H}$ decay exponentially, i.e., $\tilde{t}_n \sim e^{-|n|/\xi}$. In this \textbf{quasi-local limit}, the analytical proofs in Section~\hyperref[sec: S2]{S2} hold strictly. Figure~\ref{Fig_S1} presents the numerical verification for this baseline case ($t_3=0$).

Figure~\ref{Fig_S1}(a) shows the failure of the PBC-based entanglement spectrum in the point-gapped regime ($0.6 < t_1 < 1.4$), where the NHSE invalidates Bloch band theory. In contrast, Fig.~\ref{Fig_S1}(b) demonstrates that the non-Bloch quantities derived from $\tilde{H}$ are perfectly robust. The non-Bloch polarization $P_\beta$ (yellow triangles) and entanglement polarization $\chi(\tilde{H})$ (blue circles) coincide exactly across the phase transition at $t_1 \approx 1.078$, confirming the identity $P_\beta \equiv \chi(\tilde{H})$.

Figure~\ref{Fig_S1}(c) illustrates the circular GBZ for $t_1 = 0.8$, which guarantees the quasi-locality of $\tilde{H}$. Figure~\ref{Fig_S1}(d) confirms the physical boundary localization of the entanglement edge modes.

In contrast, the generic case $t_3 \neq 0$ (discussed in the main text) leads to a non-circular GBZ. This geometric deformation introduces non-analyticity that generates power-law decaying hoppings in $\tilde{H}$, scaling as $\tilde{t}_n \sim |n|^{-\alpha}$. This renders the system formally non-local. As analyzed in Section~\hyperref[sec: S3]{S3}, while this non-locality invalidates the Resta polarization due to the divergence of position variance, our numerical results in the main text confirm that the entanglement polarization $\chi(\tilde{H})$ remains robust, protected by the Toeplitz index.

\section{S6. Unified Description of Entanglement Spectrum and Anomalous Edge Modes}
\label{sec: S6}´

\begin{figure}[t] 
\centering
\includegraphics[width=0.8\textwidth,keepaspectratio]{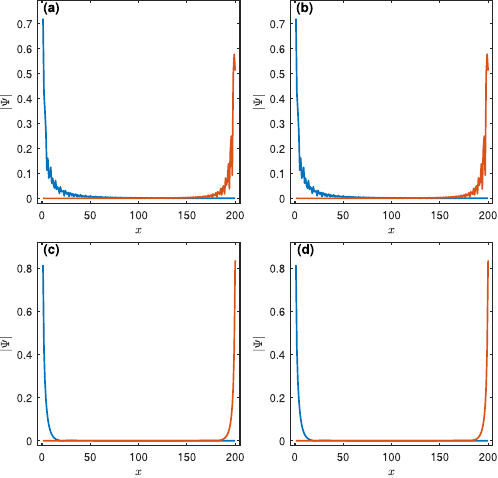}
\caption{\textbf{Chiral symmetry protection and anomalous entanglement modes.} (a)-(b) Spatial modulus of the paired entanglement modes in the non-trivial point-gapped phase ($t_1 = 0.8$). Note that the eigenvalues $\xi = 1.0057$ and $\xi' = -0.0057$ lie outside the conventional $[0, 1]$ bound, a signature of the non-Hermitian correlation matrix. Despite this, they satisfy the sum rule $\xi + \xi' = 1$, cancelling their contribution to the polarization modulo 1. (c)-(d) Spatial modulus of paired modes in the trivial phase ($t_1 = 1.5$) with conventional eigenvalues $\xi, \xi' \in [0, 1]$. The comparison demonstrates that boundary localization alone is insufficient to diagnose topology; the spectral pairing structure is the decisive factor. Parameters: $t_2 = 1, \gamma = 0.4, t_3 = 0.2, L = 800$.}
\label{Fig_S2}
\end{figure}

Conventional studies often distinguish between line-gapped and point-gapped phases, with the prevailing view that the correspondence between the entanglement spectrum (EOS) and bulk topology breaks down in the point-gapped regime. A key innovation of our study is the systematic incorporation of GBZ theory to resolve this. By constructing the biorthogonal correlation matrix $C_A$ from the reduced density matrix $\rho_A = \mathrm{Tr}_B(\vert\tilde{\Psi}^R\rangle\langle\tilde{\Psi}^L\vert)$ of the quasi-reciprocal Hamiltonian $\tilde{H}$, we demonstrate that the topological characterization via the EOS is robustly preserved even in point-gapped phases.

The correlation matrix $C_A$ is generically non-Hermitian. Consequently, its entanglement spectrum $\{\xi_\mu\}$ is complex and not restricted to the conventional probabilistic interval $[0, 1]$. Crucially, however, the underlying chiral symmetry of the quasi-reciprocal Hamiltonian $\tilde{H}$ imposes a strict pairing constraint on the EOS: for every eigenvalue $\xi_\mu$, there exists a partner eigenvalue $\xi_\nu$ such that $\xi_\nu = 1 - \xi_\mu$. This pairing structure is the mechanism that protects the topological invariant.

This protection is explicitly demonstrated in Fig.~\ref{Fig_S2}. In the point-gapped topological phase [Figs.~\ref{Fig_S2}(a) and (b)], we observe "anomalous" entanglement modes with eigenvalues $\xi = 1.0057$ and $\xi' = -0.0057$. These values violate the standard bounds of a density matrix but strictly satisfy the pairing relation $\xi + \xi' = 1$. Physically, these modes are sharply localized at opposite boundaries. In contrast, in the topologically trivial phase [Figs.~\ref{Fig_S2}(c) and (d)], the paired modes ($\xi \approx 0.96, \xi' \approx 0.04$) also exhibit boundary localization. This comparison highlights a fundamental insight: spatial localization of entanglement modes alone does not distinguish topological phases; rather, it is the specific eigenvalue structure and their sum rule that encode the topology.

This pairing structure ensures that the total entanglement polarization $\chi(\tilde H) = \sum_{\mu \in L} \xi_\mu \pmod 1$, summed over the "occupied" half of the entanglement spectrum, remains consistent with the non-Bloch bulk polarization $P_\beta$. The contributions from all paired modes—whether they are bulk bands near $\xi \approx 0/1$ or anomalous edge modes—cancel out modulo 1 ($\sum_{\text{pair}} \xi = 1 \equiv 0$). Consequently, the topological invariant is determined solely by the unpaired self-conjugate modes (e.g., those pinned at $\xi=0.5$) protected by the Toeplitz index.

Finally, we identify the emergence of boundary-localized entanglement modes with eigenvalues outside $[0, 1]$ as an intrinsic signature of the non-local regime ($t_3 \neq 0$). This phenomenon is notably absent in the quasi-local ($t_3 = 0$) limit. Crucially, these anomalous modes do not disrupt the topological calculation; due to the strict $\xi \leftrightarrow 1-\xi$ pairing, their contributions are perfectly cancelled. We propose that these modes are the entanglement-space manifestation of exceptional bound states~\cite{Lee2022a}, which emerge precisely when the effective Hamiltonian $\tilde{H}$ breaks locality, further underscoring the non-trivial interplay between non-locality and entanglement geometry.

\bibliography{references_supp.bib}